\documentclass{sig-alternate}

\newcommand\zwalkability{\mathop{\mbox{$z$-$\mathit{walkability_i}$}}}

\usepackage{balance}  
\usepackage{graphicx} 
\usepackage{times}    
\usepackage{url}      
\usepackage{amssymb,amsmath}
\usepackage{subfigure}
\usepackage[T1]{fontenc}
\usepackage[utf8]{inputenc}
\usepackage{caption}

\makeatletter
\def\url@leostyle{%
  \@ifundefined{selectfont}{\def\UrlFont{\sf}}{\def\UrlFont{\small\bf\ttfamily}}}
\makeatother
\def\pprw{8.5in}
\def\pprh{11in}

\conferenceinfo{WWW 2015,}{May 18--22, 2015, Florence, Italy.}
\copyrightetc{ACM \the\acmcopyr}
\crdata{978-1-4503-3469-3/15/05. \\
http://dx.doi.org/10.1145/2736277.2741631}

\begin{document}

\title{The Digital Life of Walkable Streets}

\numberofauthors{4}
\author{
\alignauthor
Daniele Quercia\\
       \affaddr{University of Cambridge}\\
       \email{dquercia@acm.org}
\alignauthor
 Luca Maria Aiello\\
       \affaddr{Yahoo Labs, Barcelona}\\
       \email{alucca@yahoo-inc.com}
\alignauthor
Rossano Schifanella\\
       \affaddr{University of Turin}\\
       \email{schifane@di.unito.it}
\and
\alignauthor
Adam Davies\\
       \affaddr{Walkonomics}\\
       \email{adam@walkonomics.com}
}


\setcounter{secnumdepth}{1}

\maketitle

\begin{abstract}
Walkability has many health, environmental, and economic benefits. That is why web and mobile services have been offering ways of computing walkability scores of individual street segments. Those scores are generally computed from survey data and manual counting (of even trees). However, that is costly, owing to the high time, effort, and financial costs.  To partly automate the computation of those scores, we explore the possibility of using the social media data of Flickr and Foursquare to automatically identify safe and walkable streets. We find that 
unsafe streets tend to be photographed during the day, while walkable streets are tagged with  walkability-related keywords. These results open up  practical opportunities (for, e.g., room booking services, urban route recommenders, and real-estate sites) and have theoretical implications for researchers who might  resort to the use social media data to tackle previously unanswered questions in the area of walkability.
\end{abstract}


\section{Introduction}

What makes for a good city street? Some urban planners would say the ``fabric'': the collection of streets, blocks and buildings. In ``Great Streets,'' the urbanist Alan Jacobs compared the layout of more than 40 world cities~\cite{jacobs1993great}, and found that good streets tend to have narrow lanes (making them safe from moving cars), small blocks (making them comfortable), and architecturally-rich buildings (making them interesting). Intuitively,  walking down a narrow, shop-lined street is a far safer, more comfortable, and more interesting experience than walking down an arterial between parking lots. 

Despite its importance, good street design is necessary but not sufficient for the making of great streets.  Streets, like communities, thrive on vitality~\cite{jacobs1961death}. It has been shown that the most meaningful indicator of that vitality is walkability~\cite{speck2012walkable}. This is a multi-faced concept.  Recently, in  his book ``Walkable City,'' Jeff Speck  outlines a ``General Theory of Walkability,'' identifying the four key factors that make a city attractive to pedestrians:
\begin{quotation}
``The General Theory of Walkability explains how, to be favored, a walk has to satisfy four main conditions: it must be useful, safe, comfortable, and interesting. Each of these qualities is essential and none alone is sufficient. \emph{Useful} means that most aspects of daily life are located close at hand and organized in a way that walking serves them well. \emph{Safe} means that the street has been designed to give pedestrians a fighting chance against being hit by automobiles; they must not only be safe but feel safe, which is even tougher to satisfy. \emph{Comfortable} means that buildings  shape urban streets into `outdoor living rooms', in contrast to wide-open spaces, which usually fail to attract pedestrians. \emph{Interesting} means that sidewalks are lined by unique buildings with friendly faces and that signs of humanity abound.''
 \end{quotation}
The importance of walkability goes beyond aesthetic considerations. Walkable streets not only make a city beautiful but also greatly contribute to the wealth, health, and sustainability of the city.  They contribute to wealth, not least because walkability can add 5 to 10 percent to house prices in the United States~\cite{cortright09walking,leinberger12coveted}. They contribute to health so much so that walkability is considered to be at the heart of the cure to the health-care crisis in the States by many~\cite{lee2011reversing}. Finally, they contribute to environmental sustainability. A case in point is that replacing one's light-bulbs with energy saving once a year spares as much carbon as living in a walkable neighborhood does for a week~\cite{speck2012walkable}.

The growing demand for walkable neighborhoods (especially from younger generations) has made websites  that calculate walkability (e.g., \url{walkonomics.com}, \url{walkscore.com})  popular among real estate agents, health-care agencies, and environmentalists. However, to work, those sites need to process and gather  a variety of datasets, which is financially-prohibitive.

To make walkability modeling cheap and scalable, one could resort to social media sites. That is because part of a street's vitality is, nowadays,  captured in the digital layer: street dwellers take pictures and post them on Flickr, and, when they visit places,  they share their whereabouts on Foursquare. It is therefore reasonable to assume that there might be digital footprints that distinguish walkable streets from unwalkable ones.  As a result, we study whether digital activity on Flickr and Foursquare can help us identify walkable streets in London and, more generally,  whether implicit social media data can provide walkability assessments without the need to manually collect expensive datasets.  

More specifically:
\begin{itemize}
\item We collect Flickr and Foursquare data for the 3,368 street segments in Central London (Section~\ref{sec:social_media}). One of the authors has created a web and mobile service called Walkonomics to produce safety  and walkability scores for those streets (Section~\ref{sec:walkability}). 

\item To ensure experimental validity, we review the literature and spell out four main research questions concerning safety and walkability (Section~\ref{sec:methodology}). 

\item We answer those questions upon our datasets (Section~\ref{sec:analysis}). We find that unsafe streets  tend to be photographed during the day but not at night; tend to be visited not only by males but also by females; and are identified by the presence of residential elements  of the city that have no parks. By contrast,
walkable streets are associated with residential areas and are identified by the presence of walkability-related photo tags with a correlation as high as $r=0.89$. 
\end{itemize}

Before concluding (Section~\ref{sec:conclusion}), we discuss the theoretical and practical implications of our work (Section~\ref{sec:discussion}). 

\section{Related Work}
We have heavily borrowed  from 1970s urban studies~\cite{alexander1977pattern,jacobs1993great,jacobs1961death} and from the walkability literature, most of which has been recently summarized by Jeff Speck~\cite{speck2012walkable}. Our work  is best placed within an emerging area of Computer Science research, which is often called `urban informatics.' Researchers in this area have been studying large-scale urban dynamics~\cite{crandall09mapping,cranshaw12livehoods,tassos12}, and  people's behavior when using location-based services such as Foursquare~\cite{bentley2012,cramer11performing,lindqvist11mayor}.

More closely related to this work, computational methods that automatically mine a variety of data sources to predict economic indicators have been recently developed. Eagle \emph{et al.}~\cite{eagle10} used land-line phone records to predict socio-economic indicators in English neighborhoods. More recently, to predict those indicators in London, Smith  \emph{et al.}~\cite{smith13b} used  underground transit flows. Elvidge  \emph{et al.} analyzed satellite images to extract the total surface lit during night time, and found strong correlations with  countries' Gross Domestic Product~\cite{elvidge97, elvidge01}.  Mao \emph{et al.} used mobile phone records to predict  economic indexes of ten areas of high economic activity in Cote d'Ivoire~\cite{mao13}. Traunmueller \emph{et al.} also used mobile phone records but did so to test existing urban theories (from, e.g., Jane Jacobs' work) at scale~\cite{traunmueller14mining}.

The idea of testing traditional urban theories at web scale has recently received attention. It is well-known that the layout of urban spaces plugs directly into our sense of community well-being. The $20^{th}$ century sociologist Kevin Lynch showed that everyone living in an urban environment creates their own personal ``mental map'' of the city based on features such as the routes they use and the areas they visit~\cite{lynch1960}. Lynch thus hypothesized that the more recognizable the features of a city are, the more navigable the city is. To put his theory to test, Quercia \emph{et al.} built a web game that crowdsources Londoners’ mental images of the city~\cite{quercia13maps}. They showed that areas suffering from social problems such as housing deprivation, poor living conditions, and crime are rarely present in residents’ mental images. The researchers then built another crowdsourcing game to determine which urban elements make city dwellers happy~\cite{quercia14aesthetic}. In that web game, users are shown ten pairs of urban scenes of London and, for each pair, a user needs to choose which one they consider to be most beautiful, quiet, and happy. Based on user votes, the researchers were able to rank all urban scenes according to these three attributes. By analyzing the scenes with image processing tools, they discovered that the amount of greenery in any given scene was associated with all the three attributes and that cars and fortress-like buildings were associated with sadness (they equated sadness to the low end of their `spectrum' of happiness). In contrast, public gardens and Victorian and red brick houses were associated with happiness. Upon that work, practical innovations emerged: new mapping tools that return directions that are not only short but also tend to make urban walkers happy~\cite{quercia14shortest}, and new web image ranking techniques that are able to identify memorable city pictures based on whether a neighborhood is predicted to be  beautiful or  to make people happy~\cite{Quercia2014lightweight}.

This stream of research requires access to datasets that are very difficult to get or entails the design of web engagement tools that are difficult to build. An alternative approach is to rely on more easily accessible social media data. English neighborhood deprivation has been related to Twitter topics~\cite{quercia12b} and sentiment~\cite{quercia12a}, and a new way of redefining neighborhood boundaries has been proposed upon Foursquare check-ins~\cite{cranshaw12}.

In line with this last stream of research, we propose to use user-generated content to mine street safety and walkability. In the next section, we describe the datasets, before providing the details of our methodology.

\section{Datasets}
\label{sec:social_media}

\textbf{Mapping Data.} We consider the area of Central London, which  consists of 3,368 street segments. To describe those segments, we rely on data gathered and distributed for free by OpenStreetMap (OSM) (a global group of volunteer cartographers who maintain free crowdsourced online maps) and by Ordnance Survey (the national mapping agency for Great Britain). To account for potential measurement errors when matching social media data with streets,  we add  a buffer of 22.5 meters around each street's polyline. This is common practice  and has been done using  the Vector Buffer Analysis tool provided by QGIS, a free and open-source desktop geographic information systems (GIS).

\textbf{Foursquare Data}. We collect information about all the $\sim$8K Foursquare venues in London. In Foursquare, a venue is categorized within a multi-level taxonomy. Since there are 
 hundreds of level-2 categories, categorizing venues at that level would result in a sparse dataset. To avoid that, our analysis categorizes venues with the top-level categories. That is, each venue belongs to one of these nine categories: \textit{Arts \& Entertainment}, \textit{College \& Education}, \textit{Food}, \textit{Nightlife}, \textit{Outdoors \& Recreation}, \textit{Shops}, \textit{Travel \& Transport}, \textit{Professional \& Other Places}. 

\textbf{Flickr Data}. We gather a random sample of $\sim$7M geo-referenced Flickr pictures within the bounding box of Central London. For each picture, we  summarize its popularity statistics of number of \textit{views}, \textit{favorites}, and \textit{comments}. We also collect the owner's gender and age\footnote{These were available for around 55\% of the owners in our sample.}, and the picture's both human-generated \textit{tags} (i.e., free-text annotations assigned by the photo's owner) and machine-generated tags\footnote{\url{http://www.fastcolabs.com/3037882}}. The machine-generated tags are assigned by a computer vision classifier and describe the picture's subjects (e.g., bird, tree) and  context (e.g., indoor, outdoor, night). Since we are interested in determining how many photos are taken at night on a street, we  count the number of pictures that are classified as \textit{night}, and the number of those that are classified otherwise. The machine-generated tags come with a confidence score in $[0,1]$ that reflects the probability of the tag being correclty assigned to the picture. To make sure that the photo actually is taken at night, we consider only tags that are assigned with confidence greater than $0.95$. We could have used timestamps to do the same thing, but it has been shown that they are more unreliable than considering high-confidence machine tags~\cite{thomee14time}.

\section{Walkability}
\label{sec:walkability}

\begin{figure}[t!]
\begin{center}
\includegraphics[width=0.45\textwidth]{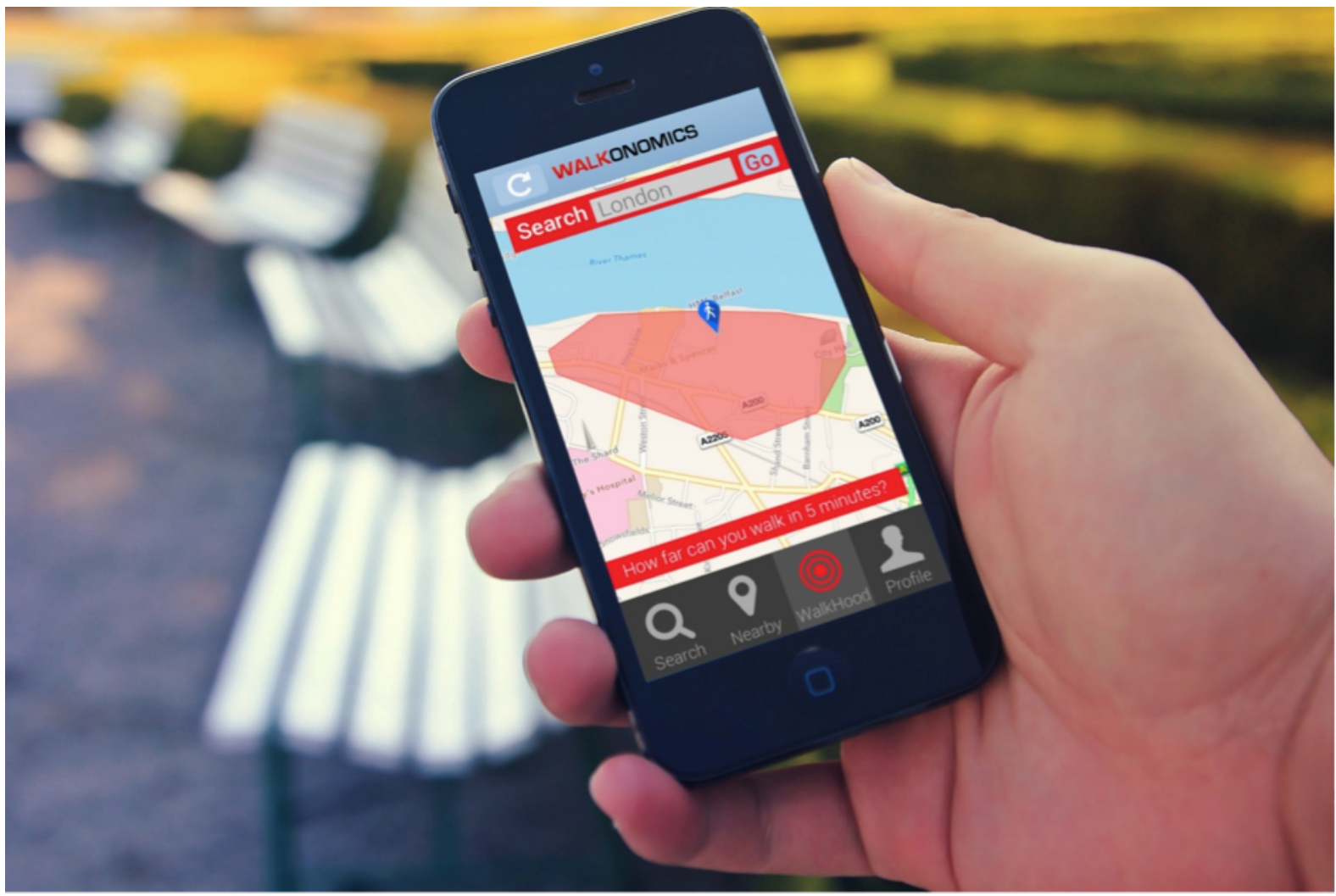}
\caption{Walkonomics app  with the \textit{``WalkHood''} feature, which shows the areas one can walk to within five minutes from current location.}
\label{fig:walkhood}
\end{center}
\end{figure}

\begin{figure*}[t!] \begin{center}
\subfigure[Walkability scores of each street segment. Green segments \mbox{ } \mbox{ } \break are very walkable, while red ones are not pedestrian-friendly.]{\includegraphics[width=.49\textwidth]{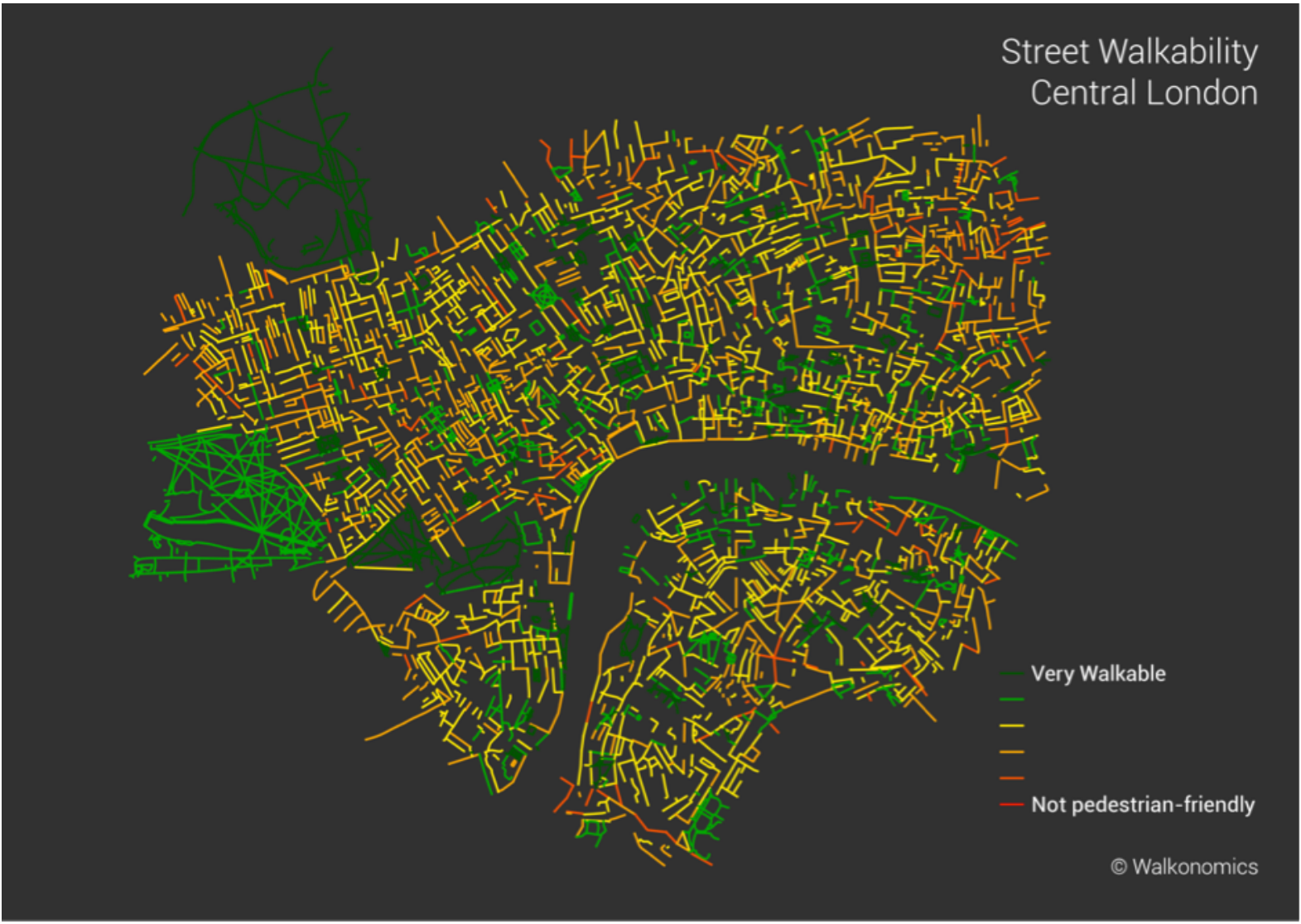} \label{fig:walkability-map}}
\subfigure[Safety from crime scores of each street segment. Green segments  have low levels of crime, while red ones have high levels.]{\includegraphics[width=.49\textwidth]{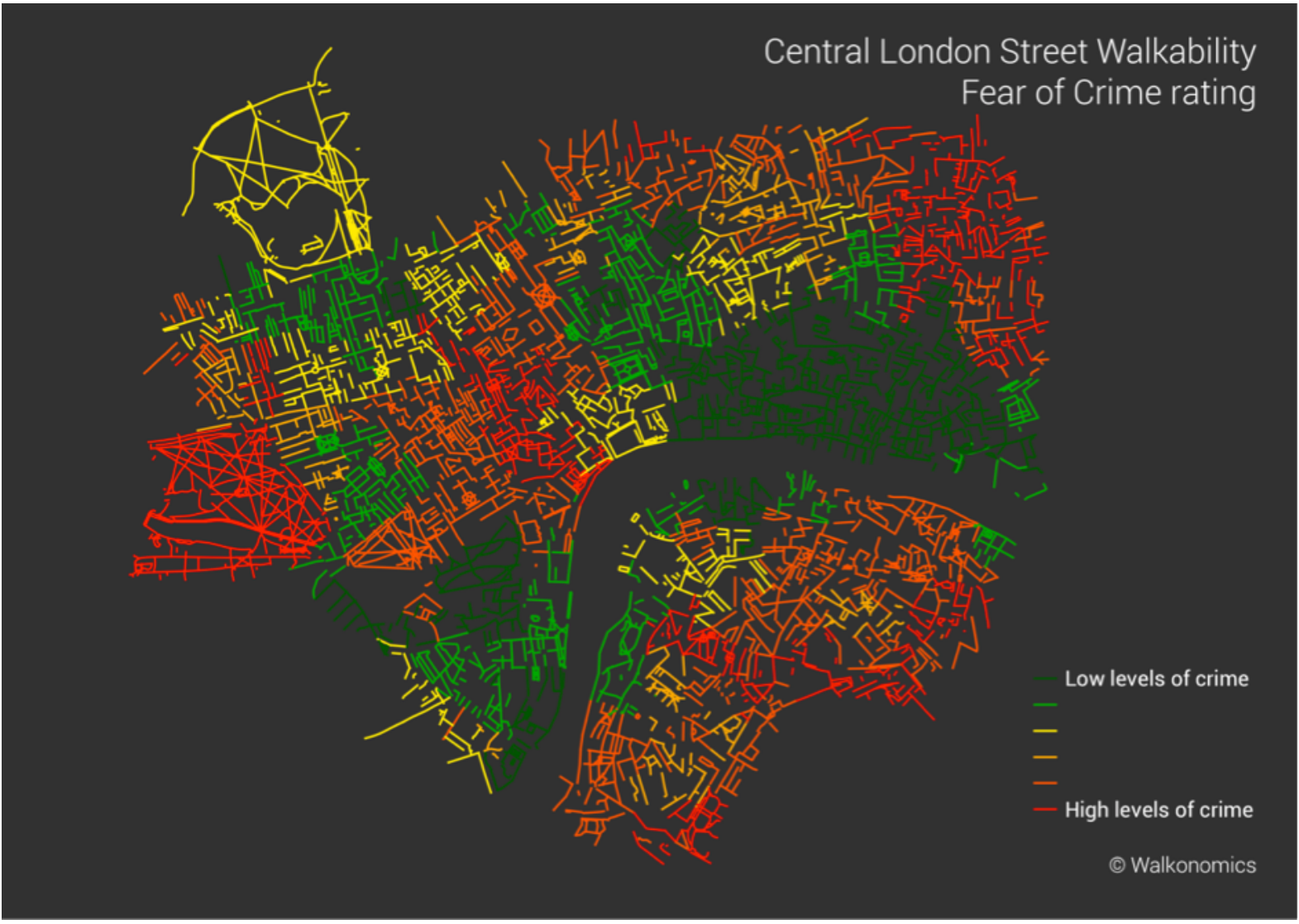} \label{fig:fearofcrime-map}}
\end{center} \vspace*{-2mm}
\caption{Maps of Central London showing to which extent each street segment is \emph{(a)} walkable, and  \emph{(b)}  safe from crime. } 
\vspace*{-1mm}\label{fig:maps} \end{figure*}

One of the authors founded Walkonomics\footnote{\url{http://www.walkonomics.com}}, a web platform and mobile app that maps and rates the pedestrian-friendliness of over 700,000 streets in England, San Francisco, Toronto and Manhattan. The mobile app has been installed in more than 8,000 devices and the website receives thousands of monthly unique visitors. Each street has five-level ratings in eight different categories. Those categories are the most important factors associated with walkability by public agencies~\cite{Methorst2010, transport2012} and existing research~\cite{ramirez-indicators, speck2012walkable}:
\begin{description}
\item \emph{Road safety.} This measures pedestrian safety from vehicle traffic. It reflects the street's type, number and severity of road accidents~\cite{transp2013}.
\item \emph{Easy to cross.} This measures how easy it is for a person to cross the street. Its score depends on the street's type (derived from OpenStreetMaps) and traffic activity. This activity is derived from the English Index of Multiple Deprivation,  which is a composite score defined at the level of census area in England (Lower Super Output Area) and is computed by the UK Office of National Statistics~\cite{depriv2010}.
\item \emph{Sidewalks.} This measures the quality and width of the street's sidewalks, and is based on the street's type.
\item \emph{Hilliness.} This measures how steep the street is. It is based on the street's slope~\cite{geological10}. 
\item \emph{Navigation.} Its score reflects the provision of pedestrian ``wayfinding'' maps and signage on the street. Location information of pedestrian signage is publicly available~\cite{Gleave14}.
\item \emph{Safety from crime.} This measures safety from street crime. This is one of the domains of the English Index of Multiple Deprivation~\cite{depriv2010}.
\item \emph{Smart and beautiful.} This measures how attractive and green the street is. It is based on the number of trees on each street, and on whether the street is in or near a park.  Information about trees and parks is extracted from OpenStreetMap.
\item \emph{Fun and relaxing.} This measures  whether a street is a fun and interesting place to be, and whether it is a relaxing environment or one dominated by vehicle traffic. Its score depends on the  number of shops, bars,  restaurants, and parks on the street (extracted from OpenStreetMap) and on the street's type.  
\end{description}
The scores for all the categories are all extracted from public data that is updated periodically but not in real time. To correct any inaccuracies or errors in assessing streets, Walkonomics allows its web and mobile phone users to upload their own street reviews. To incentivize mobile phone users to do so on the spot,  the mobile app allows them to: check the walkability of nearby streets and areas on a map; search by location, place name or post code; view search results on a map with colour-coded markers; read detailed reviews with star ratings for each category and user-generated photos; add their own ratings, reviews, photos and ideas for improvement; login using their Facebook, Twitter or email address and use their profiles to add street reviews; and see the Google StreetView of each street. The most popular feature of the app is the \textit{``WalkHood''} map (Figure \ref{fig:walkhood}). This shows a polygon of the areas a user  can walk to within 5 minutes from the current location.

\begin{figure}[t!] 
\subfigure[Walkability \break (min:1.5, median:2.5, max:4.63)]{\includegraphics[width=.23\textwidth]{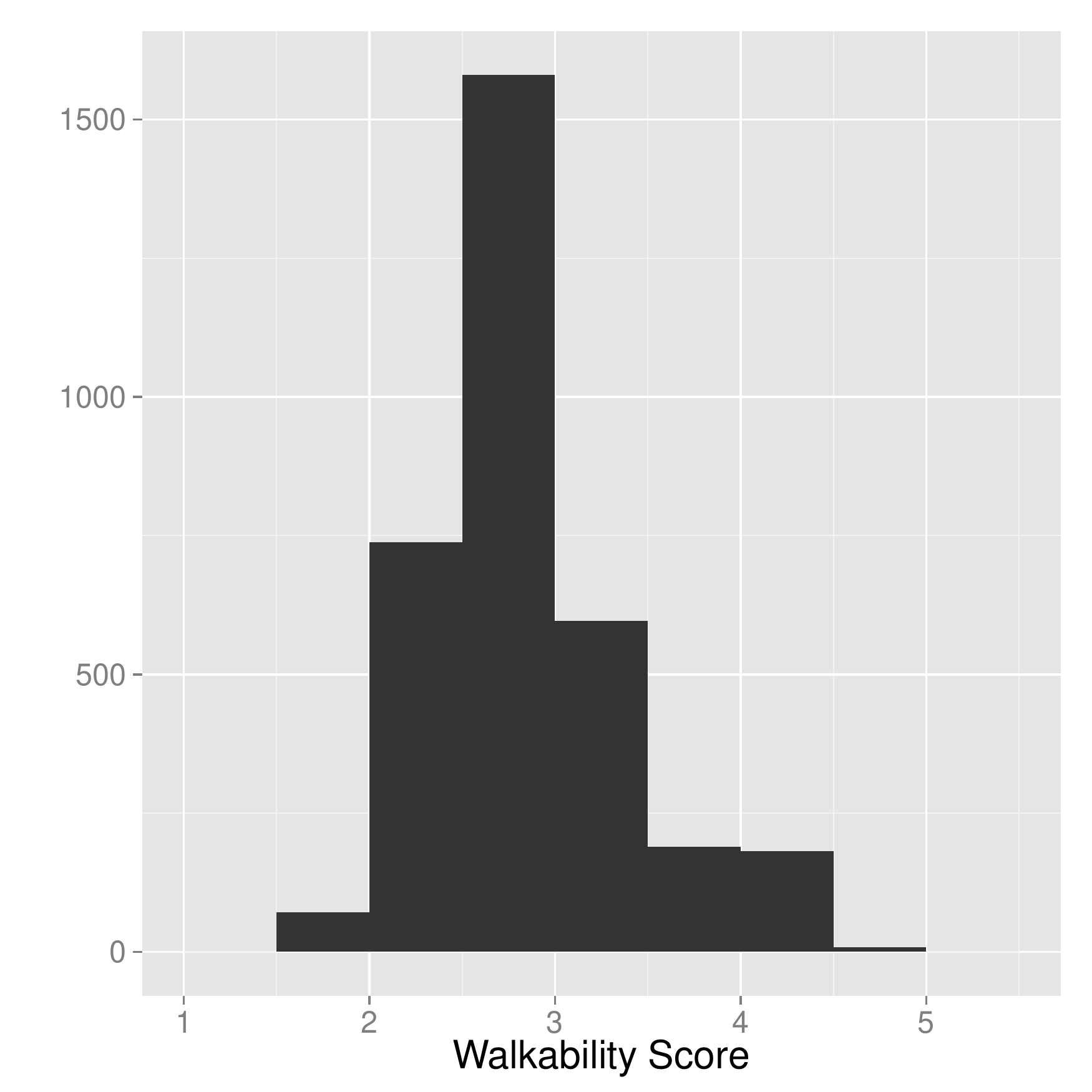} \label{fig:hist_walkability}}
\subfigure[Safety \break  (min:0.5, median:3, max:5)]{\includegraphics[width=.23\textwidth]{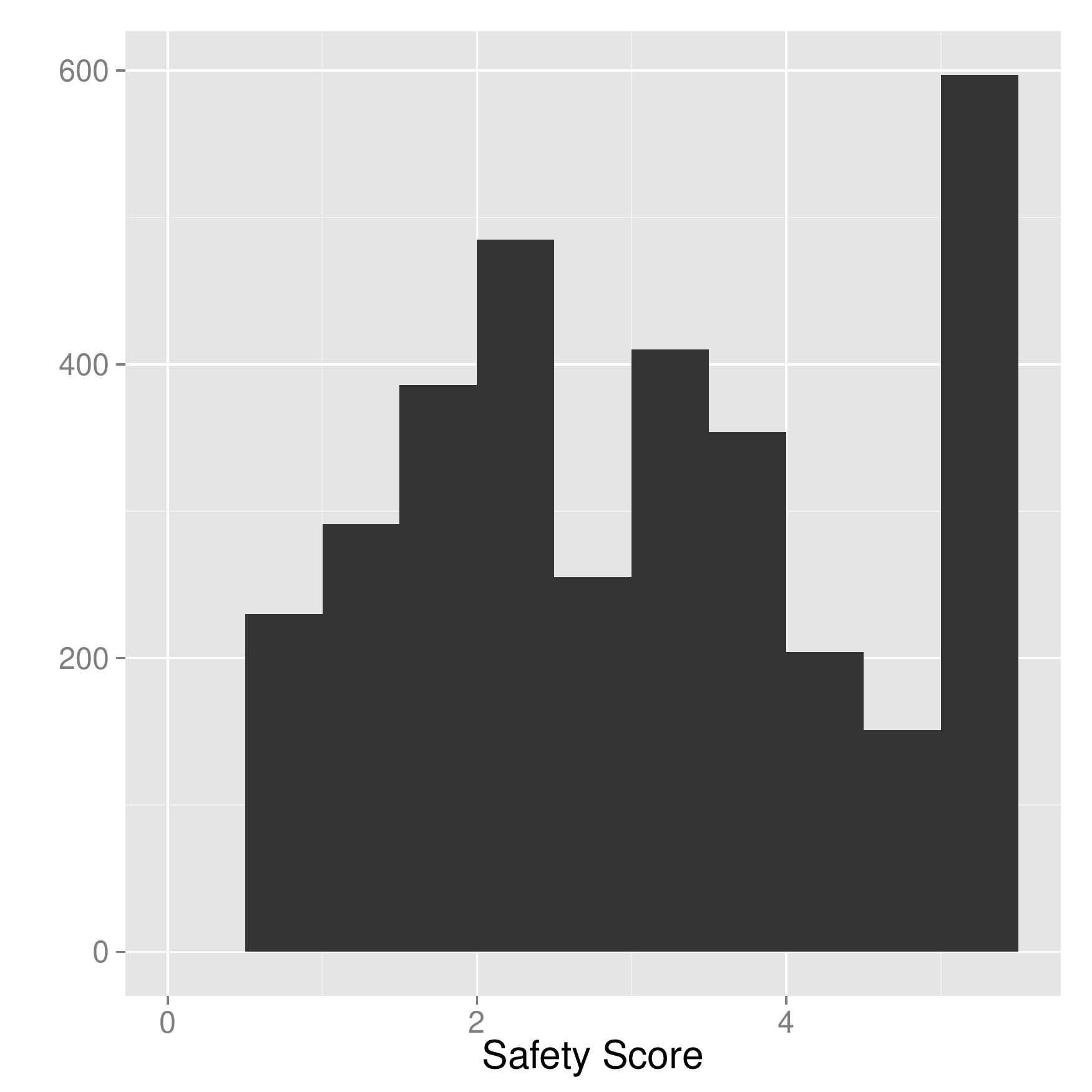} \label{fig:hist_safety}}
\caption{Frequency distributions of the walkability and safety scores at the level of street segment. The scores are defined from 1 to 5. The walkability scores in \emph{(a)} are centered around a median of 2.5, while the safety ones in \emph{(b)} are more uniformly distributed.} 
\vspace*{-1mm}\label{fig:frequency_distributions} \end{figure}

The street's overall walkability score is the average of the eight categories, equally weighted (Figure~\ref{fig:walkability-map}). Since urban crime is the dimension among those provided by UK Office of National Statistics most related to walkability, we start with a few research questions about crime (which has been widely-studied in the urban context~\cite{sampson12when}) to  then move on with questions about  walkability. To ease comparison, Figure~\ref{fig:fearofcrime-map} maps the ``safety from crime'' scores in Central London, and Figure~\ref{fig:frequency_distributions} shows the frequency distributions of walkability and safety.

\begin{figure*}[t!] 
\subfigure[Average Age \break (min:26, med:40, max:63)]{\includegraphics[width=.19\textwidth]{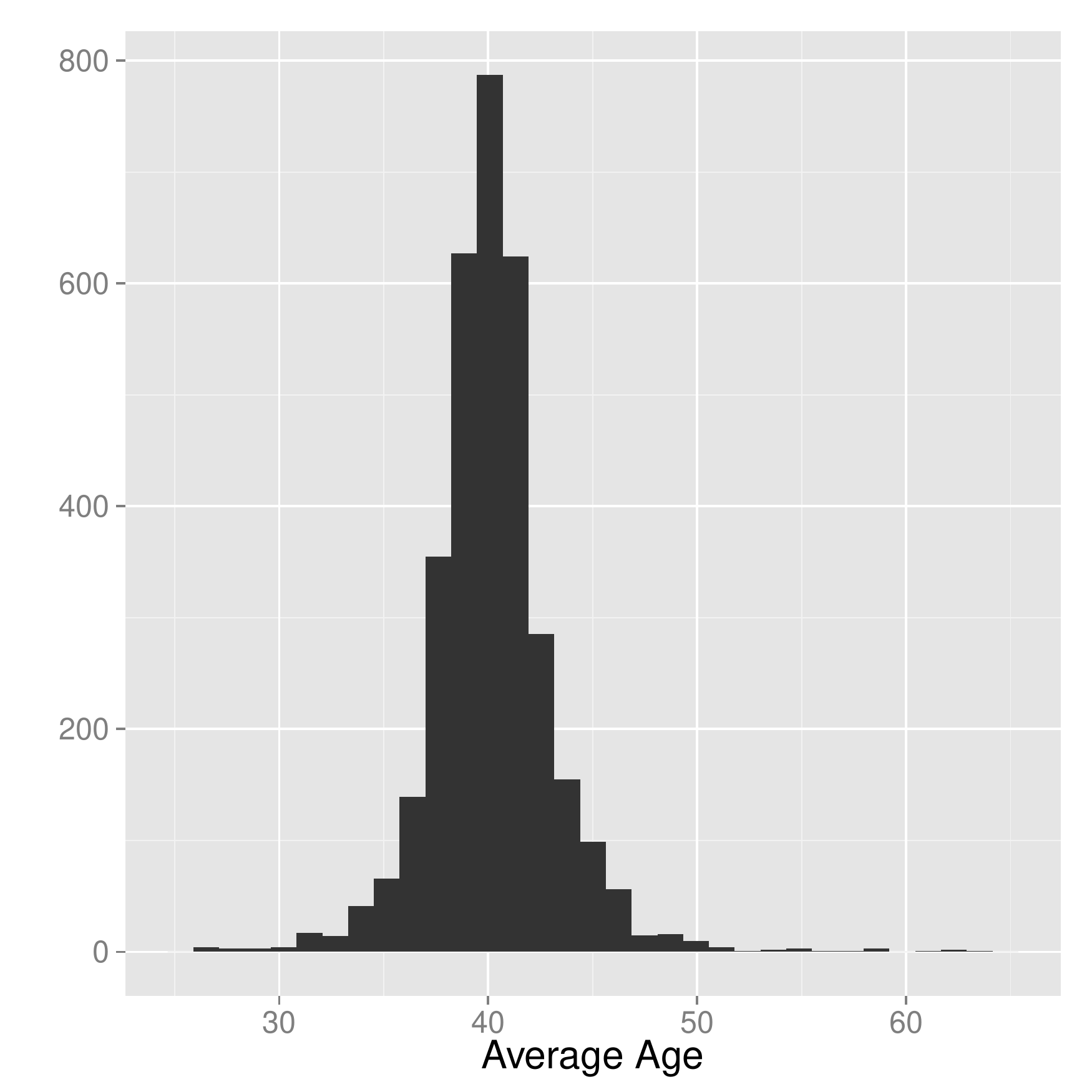} \label{fig:hist_walkability}}
\subfigure[Male \break  (min:1,med:87,max:4564)]{\includegraphics[width=.19\textwidth]{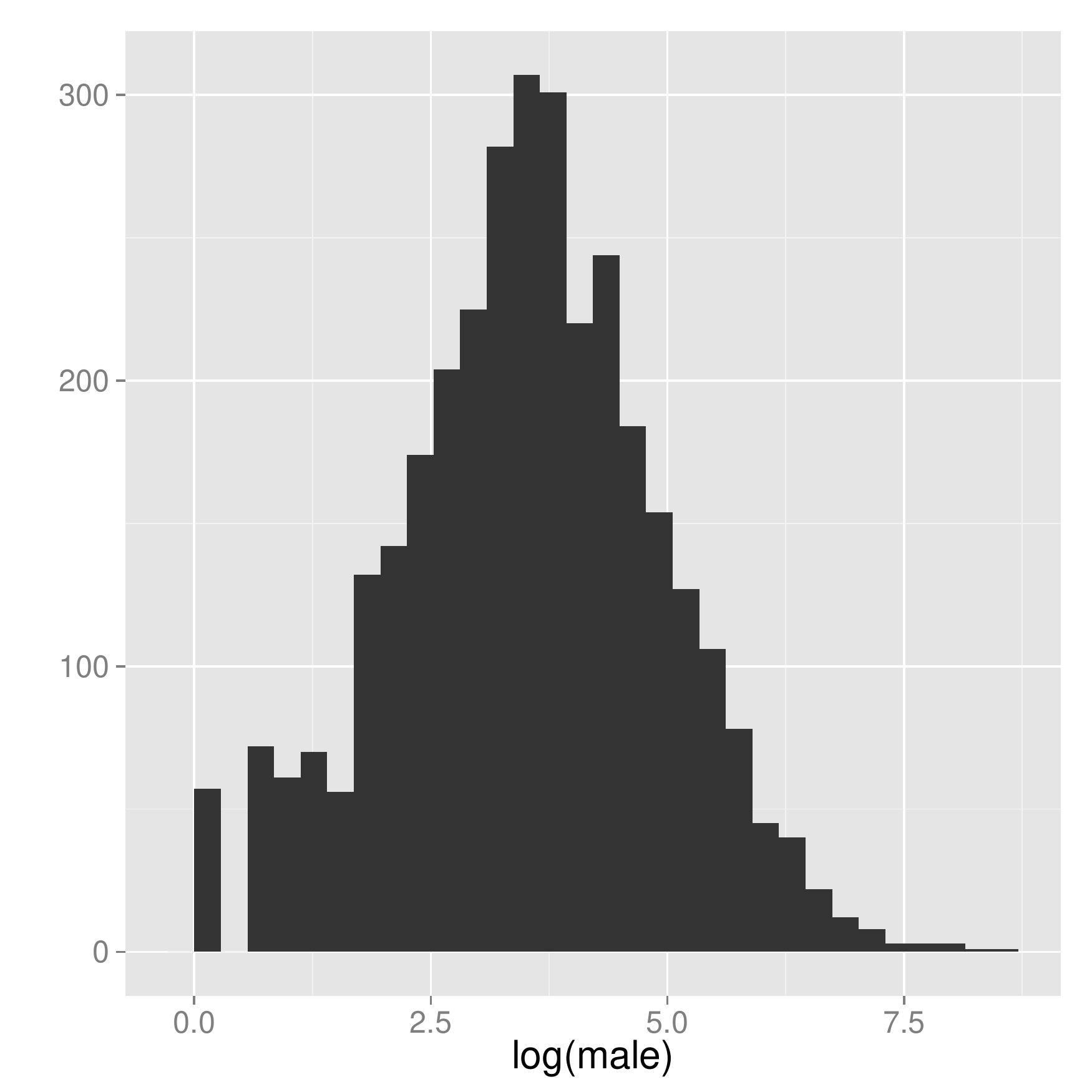} \label{fig:hist_safety}}
\subfigure[Female \break  (min:1,med:20,max:2760)]{\includegraphics[width=.19\textwidth]{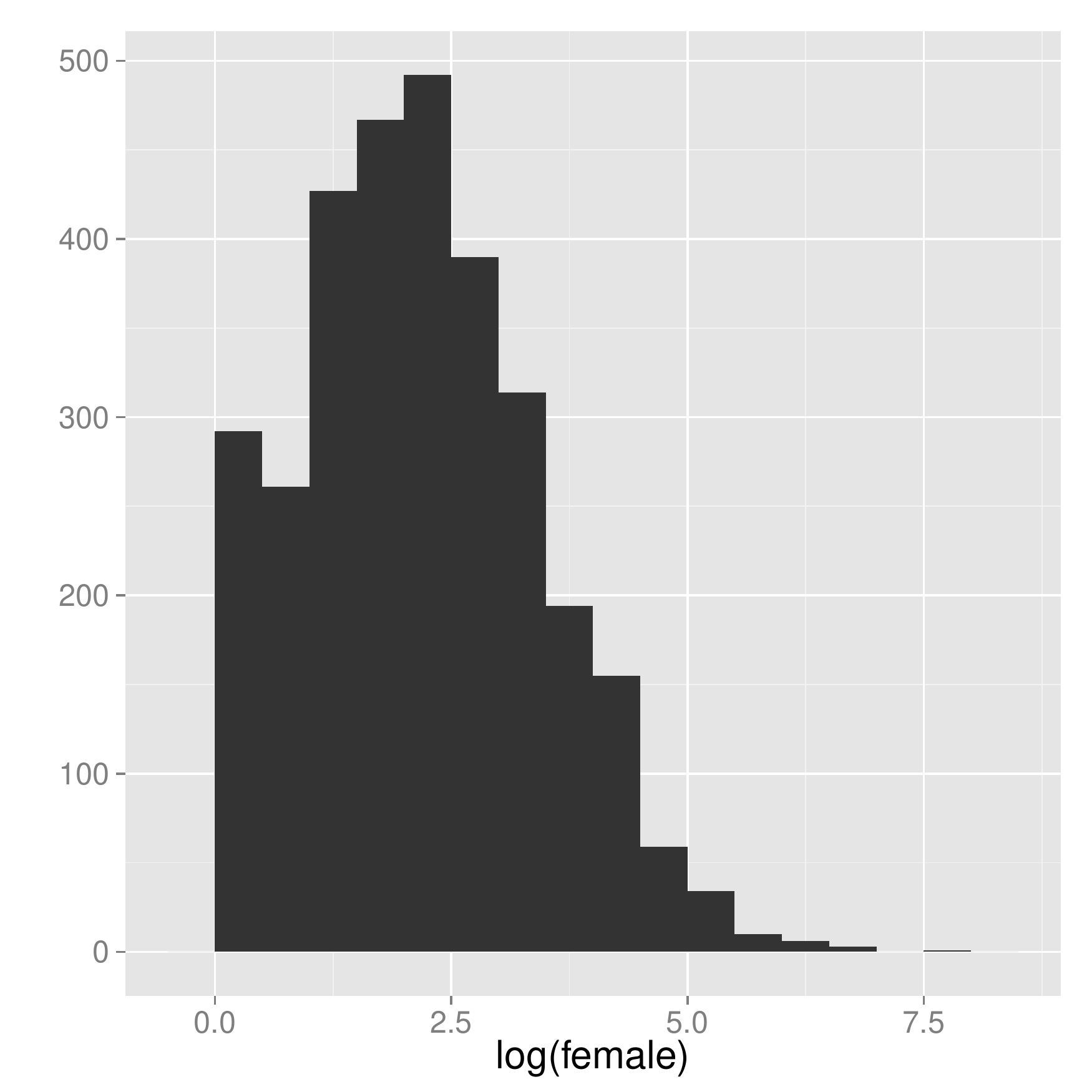} \label{fig:hist_safety}}
\subfigure[Flickr tags \break  (min:2,med:4.7K,max:968K)]{\includegraphics[width=.19\textwidth]{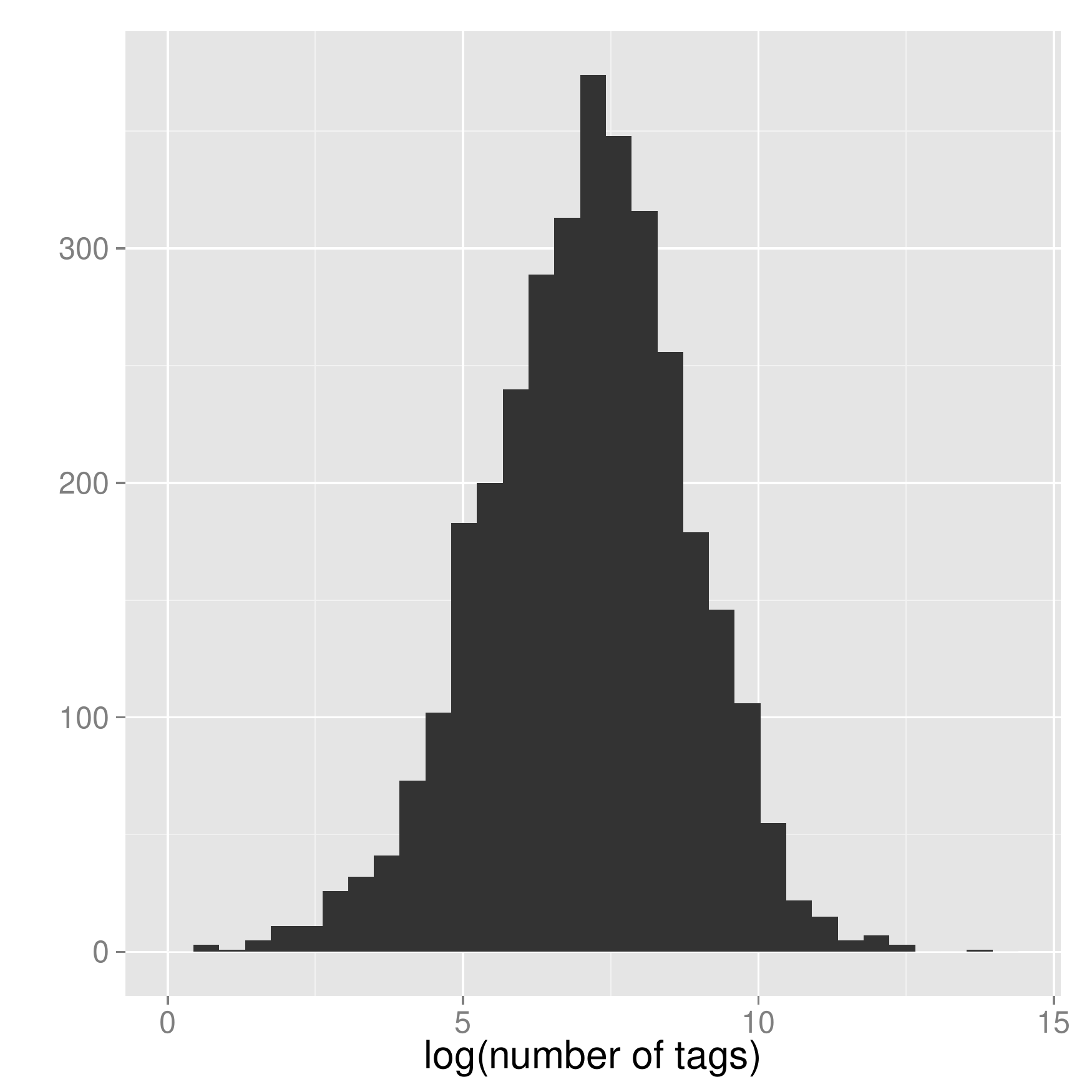} \label{fig:hist_safety}}
\subfigure[Foursquare places \break  \mbox{  } (min:1,med:2,max:27)]{\includegraphics[width=.19\textwidth]{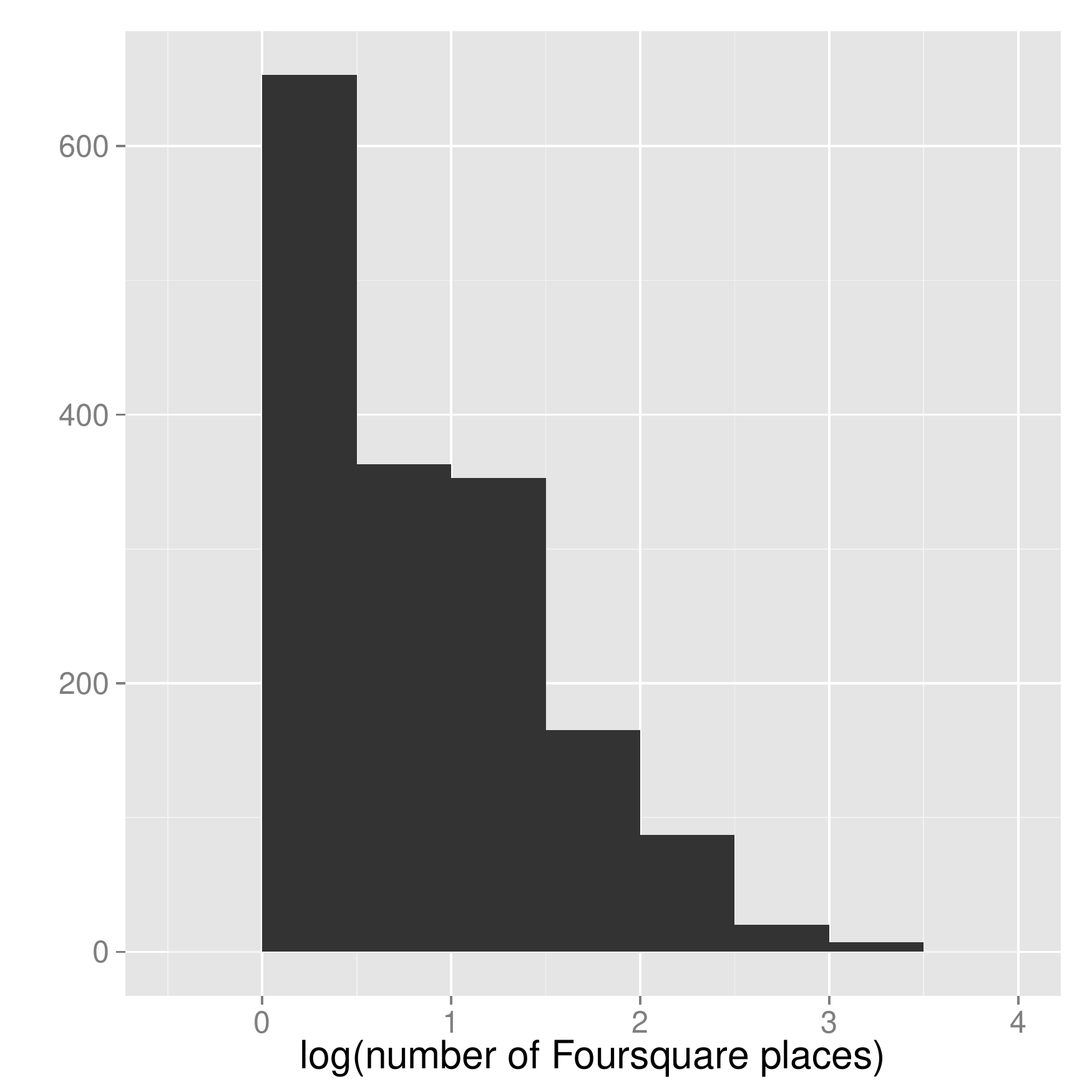} \label{fig:hist_safety}}
\caption{Frequency distributions of Flickr and Foursquare activity features. Below the plot of each feature's frequency distribution, we report the minimum value, median value, and maximum value.  All the features but age are log-transformed as the original values are skewed.} 
\vspace*{-1mm}\label{fig:hists} \end{figure*}

\section{Method}
\label{sec:methodology}

Critics might rightly say that we are not sure whether the scores we have just introduced actually measure what they are meant to measure (i.e., safety, walkability). To assess the validity of those scores, we need to theoretically derive hypotheses concerning, say, walkability (e.g., it is associated with the absence of cars) and test those hypotheses upon those scores. If the hypotheses receive support  (e.g., the absence of cars is indeed found to be empirically associated with the walkability scores), then that speaks to the validity of the scores (concurrent validity). We thus derive hypotheses concerning safety and walkability next.

\subsection{Research Questions on Safety}

In the early 1960s, Jane Jacobs explored the relationships between urban decays, social interactions, and crime. She showed that nothing is safer than a city street that everybody uses, and called this phenomenon ``the eyes on the street''~\cite{jacobs1961death}. 

In ``The Ecology of Night Life,'' Shlomo Angel indeed showed that areas of very low or very high pedestrian density suffer from much less crime~\cite{angel68}. ``At night, street crimes are most prevalent in places where there are too few pedestrians to provide natural surveillance, but enough pedestrians to make it worth a thief's \break while''~\cite{alexander1977pattern}. Based on that, we posit our first research question:
\begin{description}
\item[R1:] Can safe streets be identified by night activity?
\end{description}
In a similar vein, one could consider gender differences, in that, streets that men use might differ from those that women use in terms of safety from crime.  However, it is unclear the nature of this relationship.  One might hypothesize that safe streets are used by men and women alike, and unsafe ones are used by men only (women are likely to shy away). But one might also hypothesize the opposite:  ``to make it worth a thief's while'' (as Alexander puts it),  unsafe streets are so because they are predominantly used by women. Similar considerations go for age -- streets that younger adults use might differ from those that older ones use in terms of safety. All this leads to our second research question:
\begin{description}
\item[R2:] Can safe streets be identified by activity segmented by gender or age?
\end{description}

Jacobs' ideas about urban decays led to what urbanists now call ``crime prevention through environmental design''~\cite{hess2008introduction}. This is based on the premise that the physical environment can be designed or manipulated to reduce fear of crime. One of the key strategies for crime prevention is activity support. The idea is that encouraging legitimate activity in public places (e.g., a basketball court, community center) helps discourage crime~\cite{ceccato2012urban}.  Therefore one expects that a safe street would offer places that encourage legitimate activity. Hence, our third research question:
\begin{description}
\item[R3:] Can safe streets be identified by the presence of specific types of places?
\end{description}

\subsection{Research Questions on Walkability}
Recall that, in Jeff Speck's General Theory of Walkability, a walk has to satisfy four main conditions. It must be not only safe, comfortable, and interesting, but also useful~\cite{speck2012walkable}. By useful, he means that ``most aspects of daily life are located close at hand.'' The most widely-used (albeit oversimplified) definition of walkability indeed concerns access to opportunities: the more miles one has to travel  from a place  for daily errands, the less walkable's the place~\cite{ball2012livable}. This begs our next research question:

\begin{description}
\item[R4:] Can walkable streets be identified by the presence of specific types of places?
\end{description}

The concept of walkability goes beyond the idea of access to opportunities though. To partly capture this richness, we gather the literature on walkability to produce a list of walkability-related keywords. With such a list, we aim at answering our final research question:

\begin{description}
\item[R5:] Can walkable streets be identified by walkability-related photo tags?
\end{description}

The frequency distributions of the activity features we will use to answer those questions are summarized in Figure~\ref{fig:hists}.

\section{Analysis}
\label{sec:analysis}

To answer the five research questions, we need to derive suitable Flickr and Foursquare activity features. However, before doing so, we need to ascertain whether those activity features are reliable. Without reliable measures of night activity on Flickr, of the presence of specific Foursquare places, and of the presence of Flickr photo tags, we cannot  test our hypotheses.  In general, there are three main types of error that reduce reliability: measurement error, specification error, and sampling error. To minimize the error that inevitably occurs in measuring Flickr and Foursquare activity (measurement error), we borrow measurement procedures from the literature~\cite{Cheng:EtAl:11,thomee14time}.  To minimize the effect of Flickr and Foursquare biases (e.g., Flickr pictures are  taken predominantly during the day and by men),  we borrow normalization measures (e.g., $z$-transformations) from previous studies~\cite{kramer10unobtrusive}. Finally, to partly generalize our measurements to users not in our sample (sampling error), we will determine the minimum amount of data at the street level (e.g., number of photos per street) required to have measurements yielding the same results on repeated trials.

\subsection{Research Question 1}
\emph{Can safe streets be identified by night activity?}

\begin{figure*}[t!] \begin{center}
\subfigure[Average street safety for street segments grouped by their \mbox{    } \emph{photo@night} scores in three bins. Safety increases for streets \mbox{ } \break that are increasingly photographed at night. Whiskers represent the \mbox{ }  $2^{nd}$ and $98^{th}$ percentiles.]{\includegraphics[width=.50\textwidth]{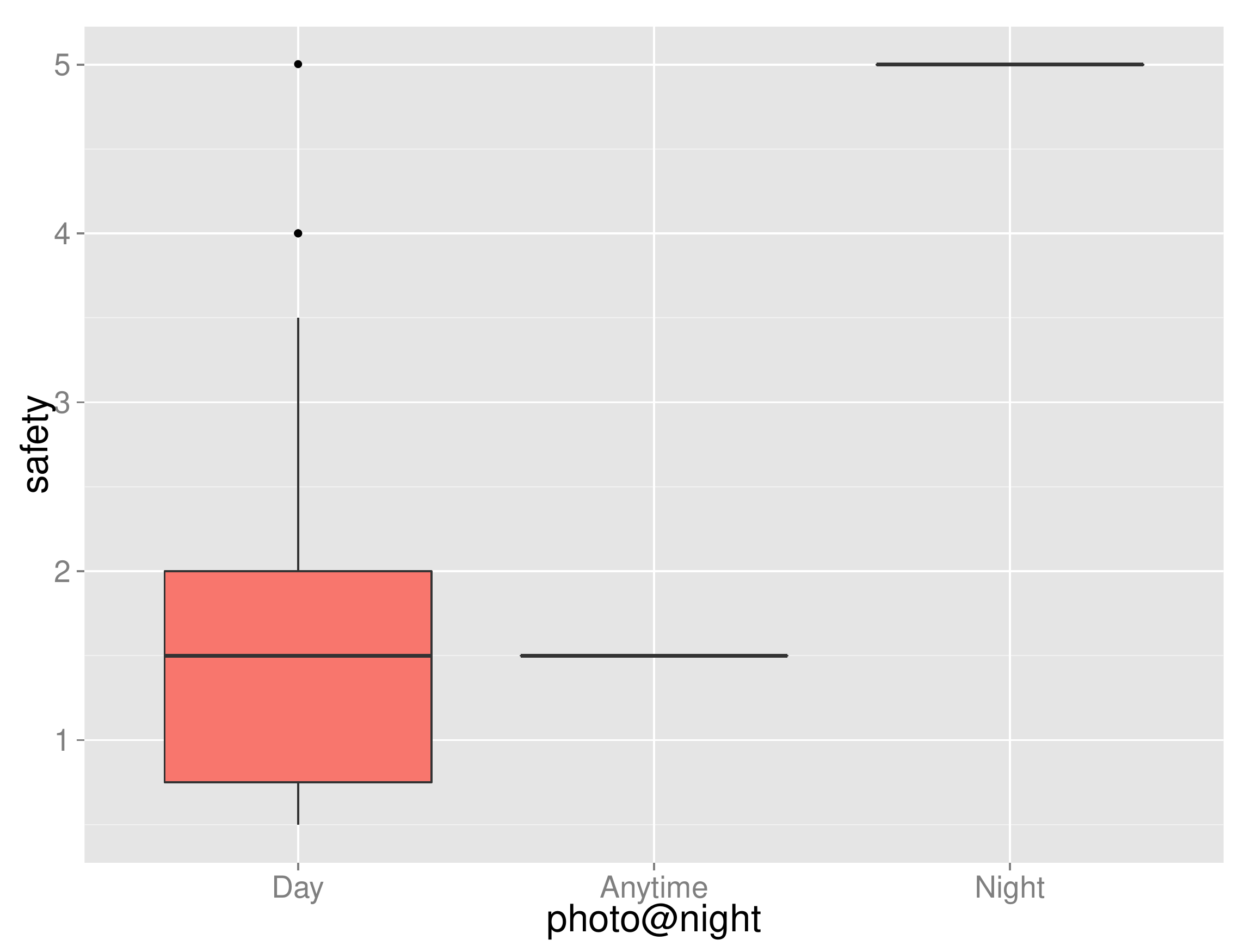} \label{fig:safe_night}}
\subfigure[Pearson correlation coefficient between street safety and  \emph{photo@night} as the number of photos on each segment increases. The shaded area indicates the number of photos per street segment after which the correlation becomes stable. ]{\includegraphics[width=.38\textwidth]{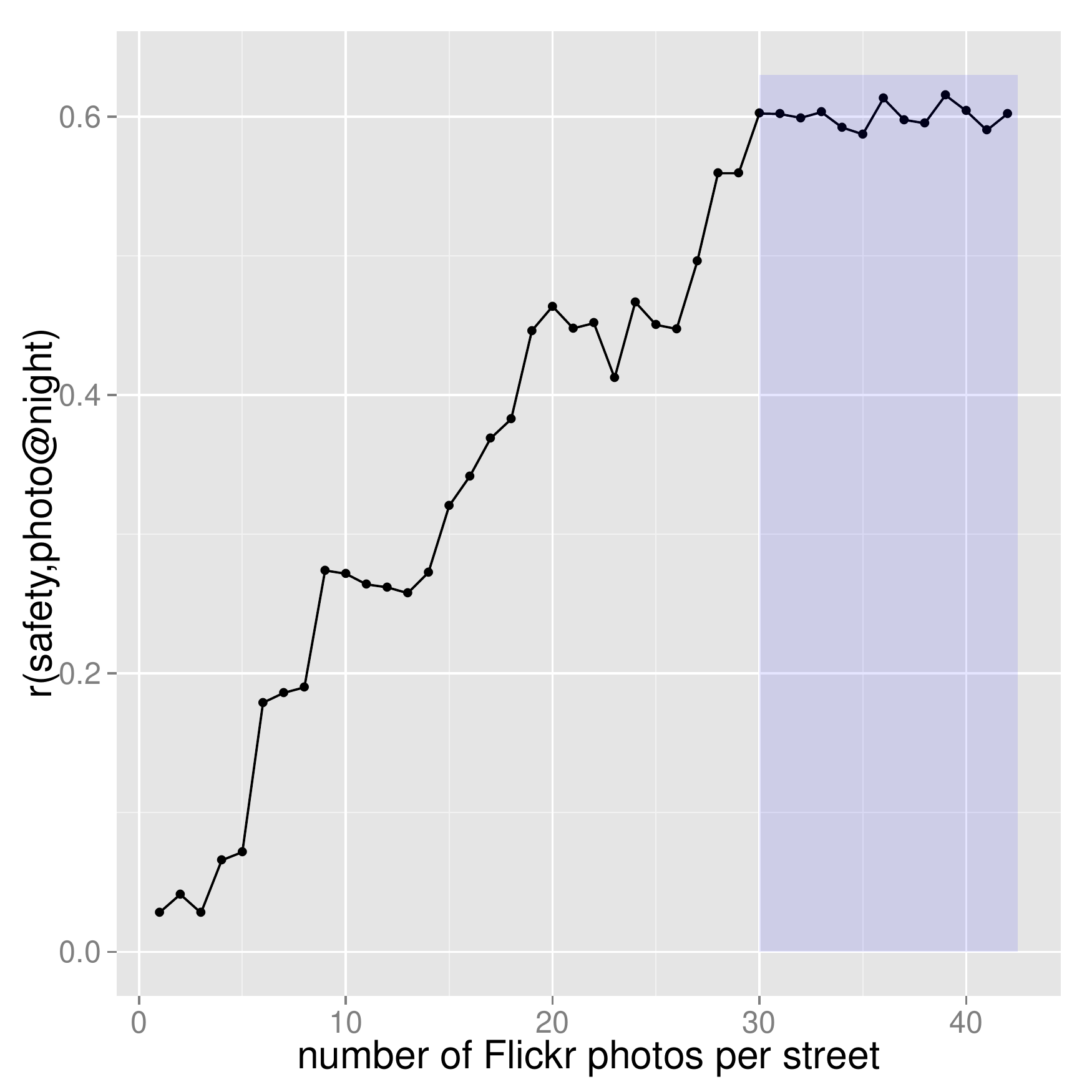} \label{how_many_night_crime}}
\end{center} \vspace*{-2mm}
\caption{The digital life of safe streets: night activity. Safe streets tend to be photographed at night as well.} 
\vspace*{-1mm}\label{fig:prob_all_numinvestors} \end{figure*}

For each street segment $i$, we compute a \emph{photo@night} score :
\begin{equation*}
photo@night_i = \frac{n_i - \mu_n}{\sigma_n} - \frac{o_i - \mu_o}{\sigma_o},
\end{equation*}
where $n_i$ ($o_i$) is the fraction of pictures taken at night (not at night) on street segment $i$; $\mu_m$  ($\mu_o$) is the fraction of night (not night) pictures, averaged across all segments; and $\sigma_n$ ($\sigma_o$) is the corresponding standard deviation. The resulting measurement is the $z$-score of the fraction of night pictures and  accounts for the unbalances  of pictures taken at night \emph{vs.} day\footnote{On Flickr, pictures are taken more during the day than at night.}.

Having each street's score at hand, we can now correlate it with \emph{safety from crime}. In so doing, we learn a strong positive correlation of $r=0.60$:  safe streets are photographed not only during the day but also at night, while unsafe ones mostly during the day. To further validate this statement, we group streets by their  \emph{photo@night} scores and test whether streets with higher scores are, on average, safer. By grouping streets into three bins,  we find clear-cut evidence (Figure~\ref{fig:safe_night}): streets in the first bin (those photographed during the day) are far less safe (with a median fear of crime of 1.4) than streets in the last bin (those photographed mainly at night).

One might now wonder whether those results are observed only for Flickr-data-rich streets. To test that, we see how the previous   correlation between safety and \emph{photo@night} changes depending on the number of Flickr photos on each street. As one expects, it does change: the more photos, the higher the correlations. However, the amount of data needed to have a stable correlation is limited: aggregating all the streets with at least 30 photos results into stable correlations of $r > 0.6$ (Figure~\ref{how_many_night_crime}). That number of photos is extremely low considering that the mean number of photos per street segment is  832, and the maximum goes up to 131K.

\subsection{Research Question 2}
\emph{Can safe streets be identified by activity segmented by gender or age?}

\begin{figure*}[t!] \begin{center}
\subfigure[Average safety score for street \mbox{ } \mbox{ }  \mbox{ } \break segments grouped by  whether their  \mbox{ } \mbox{ }  \mbox{ } \break manhood scores are in the lower  \mbox{ } \mbox{ }  \mbox{ } \break quartile (Q1), second quartile (Q2),  \mbox{ } \mbox{ }  \mbox{ } \break upper quartile (Q3), and interquartile  \mbox{ } \mbox{ }  \mbox{ } \break range (IQR). Whiskers represent the $2^{nd}$ \mbox{ } \mbox{ } and $98^{th}$ percentiles]{\includegraphics[width=.32\textwidth]{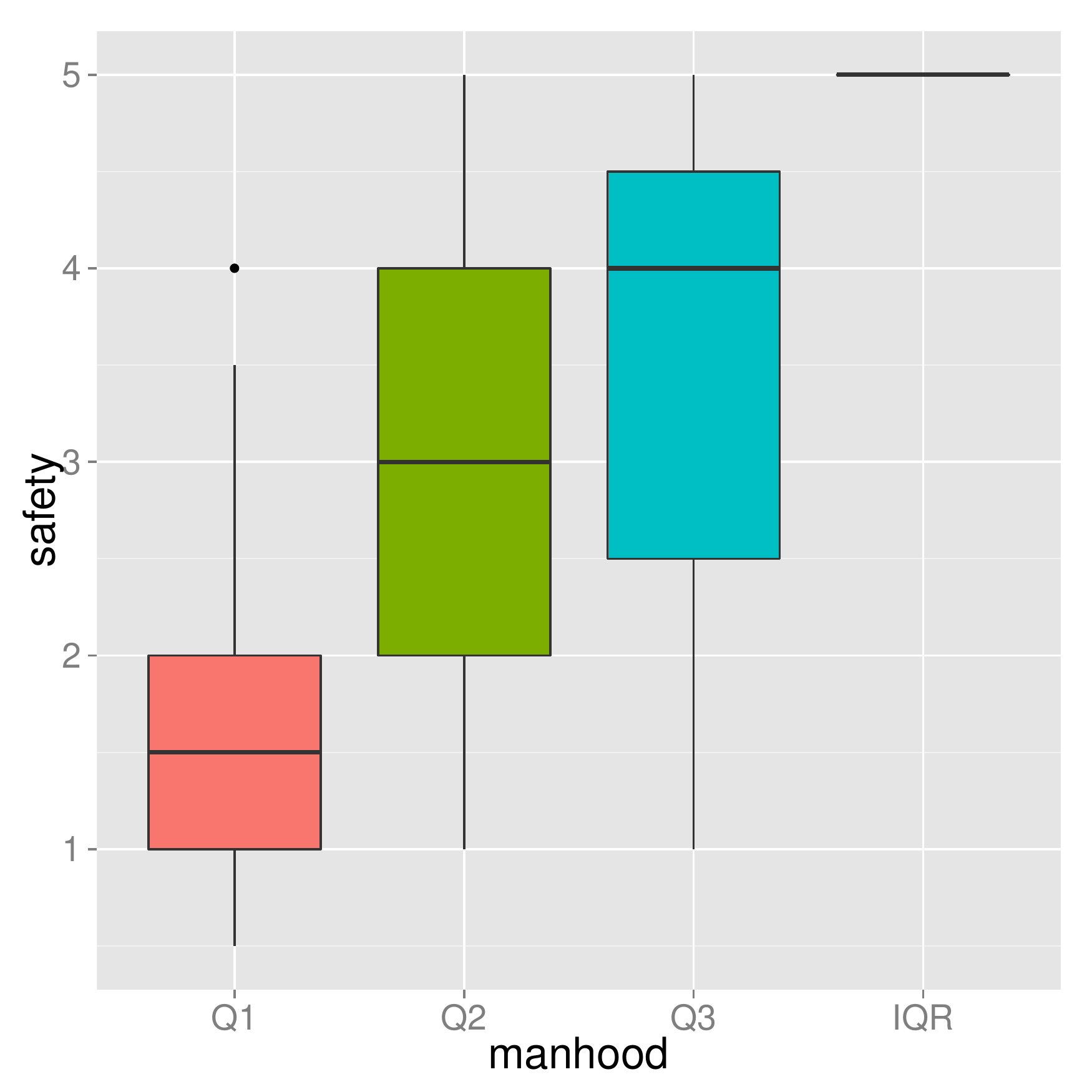} \label{fig:crime_male}}
\subfigure[Correlation coefficient  \mbox{ } \mbox{ }  \mbox{ } \break $r$(safety, street's manhood)  \mbox{ } \mbox{ }  \mbox{ } \break for segments of differing number of  \mbox{ } \break users. The shaded area indicates the \mbox{ } \break number of users per street segment \mbox{ } \break after which the correlation becomes stable. ]{\includegraphics[width=.32\textwidth]{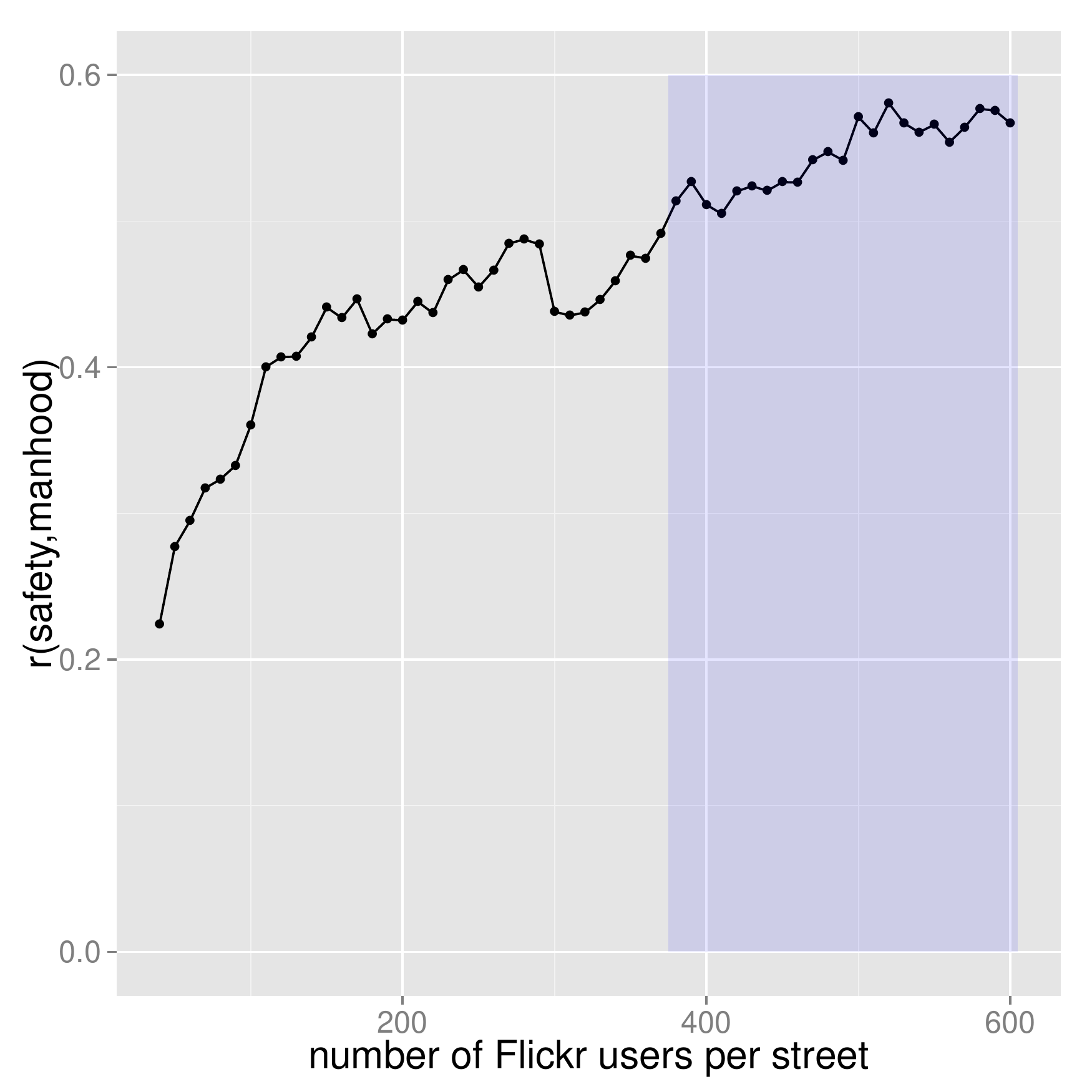} \label{how_many_manhood_crime}}
\subfigure[Correlation coefficient  \mbox{ } \mbox{ }  \mbox{ } \break $r$(safety,  \mbox{   } dwellers' average age)  \mbox{ } \mbox{ }  \mbox{ } \break for segments of differing number of users. The shaded area indicates the number of users per street segment after which the correlation becomes stable. ]{\includegraphics[width=.32\textwidth]{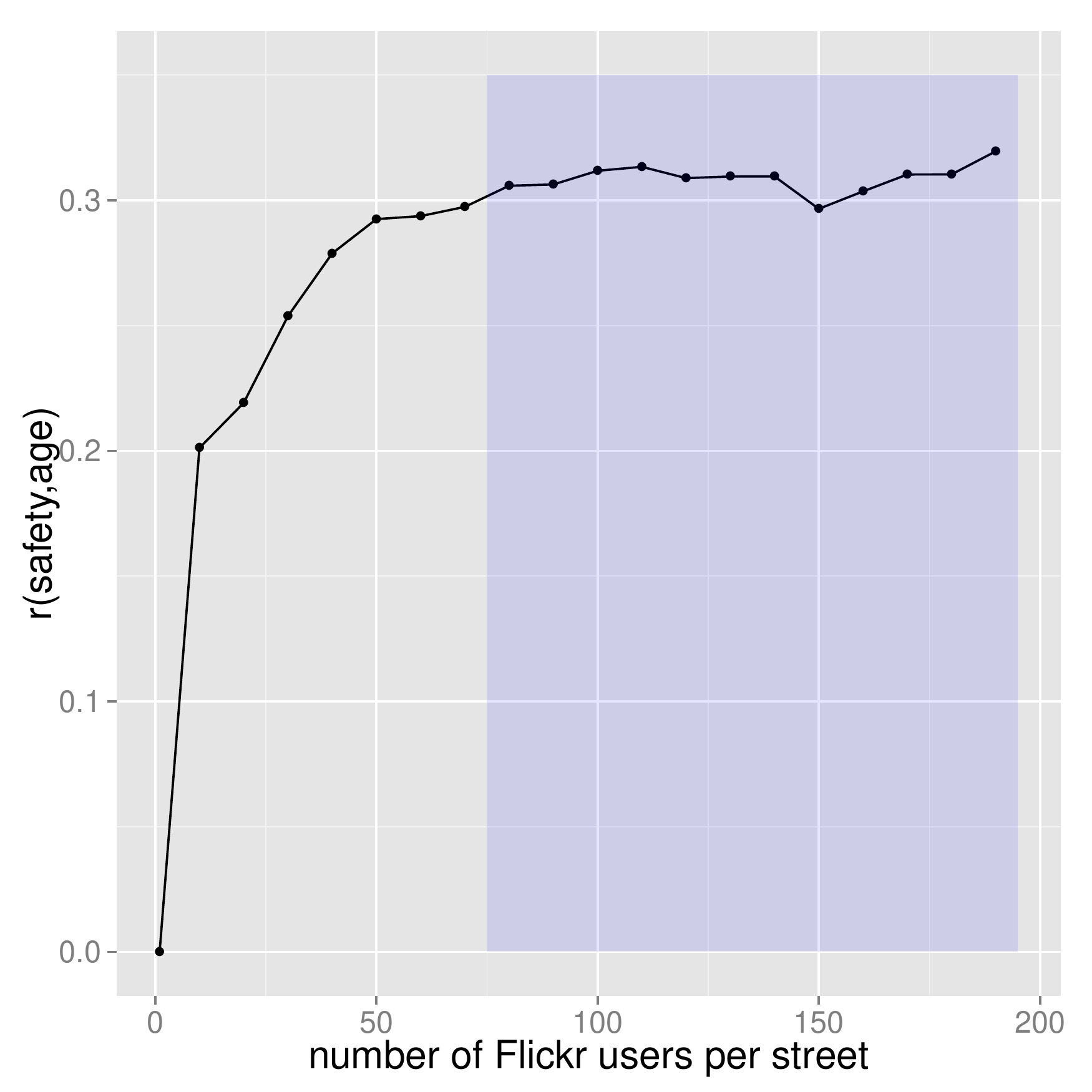} \label{how_many_age_crime}}
\end{center} \vspace*{-2mm}
\caption{The digital life of safe streets: gender and age. Safe streets tend to be increasingly photographed by men.} 
\vspace*{-1mm}\label{fig:prob_all_numinvestors} \end{figure*}

For each street segment $i$, we compute a ``manhood'' score:
\begin{equation*}
manhood_i = \frac{m_i - \mu_m}{\sigma_m} - \frac{f_i - \mu_f}{\sigma_f},
\end{equation*}
where $m_i$ ($f_i$) is the fraction of male (female) users who have taken a picture on street segment $i$; $\mu_m$  ($\mu_f$) is the fraction of male (female) users, averaged across all segments; and $\sigma_m$ ($\sigma_f$) is the corresponding standard deviation. This is the $z$-score of the fraction of male users normalized to  account for the unbalanced distribution of male and female users on Flickr. 

By correlating manhood with safety (from crime), we find a positive correlation of $r=0.58$, suggesting that safe streets tend  to be  visited by a predominantly male population. This parallels Alexander's suggestion that crime focuses on areas in which there are enough victims ``to make it worth a thief's while''~\cite{alexander1977pattern}. To further validate this finding, we group streets by their male scores and test whether streets with higher scores show, on average, higher safety. By binning streets into quartiles,  that is exactly what we find (Figure~\ref{fig:crime_male}): streets in the lower quartile (those photographed more by females than males) are unsafer (with a median safety of 1.4) than streets in the last quartile (with a median of 4).

Our second hypothesized relationship for safety is that with dwellers' age. In our sample, users have a median age of 40 and are in the range [26,63] (Figure~\ref{fig:hist_walkability}). By averaging the age of users who took pictures on each street, we indeed find a positive correlation with safety ($r=0.32$). The same correlation holds for median age. 

To test whether those results are observed only for Flickr-data-rich streets, we see how the previous two correlations safety-manhood and safety-age change for streets that differ from the number of Flickr users  they have. As one expects, the correlations do change (i.e., the more users, the higher the correlations) but it does not require many users to become stable: safety-manhood correlations become stable ($r > 0.5$) after collecting the gender of at least 380 users (Figure~\ref{how_many_manhood_crime}), and safety-age ones become stable ($r > 0.3$) after collecting the age information for only 80 users (Figure~\ref{how_many_age_crime}).

\mbox{ } \\

\subsection{Research Question 3}
\emph{Can safe streets be identified by the presence of specific types of places?}

\setlength\tabcolsep{1mm}
\begin{table}[t!]
\begin{center}
{
\begin{tabular}{lrrl} \hline
Predictive Variable & $\beta$ (walkability) & $\beta$ (safety) \\ \hline
Outdoors & 1.701 & 16.543  \\
Arts & 6.303* & -13.036* \\
College & -4.812 & 13.820   \\
Food & 0.161 & 2.380   \\
Nightlife & -8.947 & -9.897*  \\
Work & 5.282* & 8.731** \\
Residential & 21.290** & -60.628  \\
Shopping & -1.195 & -0.370  \\
\hline
\end{tabular} }
\end{center}
\caption{The predictive variables in the two linear models for walkability (column 2) and safety (column 3). Significance:  ** $p<0.001$, * $p<0.01$, . $p<0.05$.}\vspace{0.2cm}
\label{table:predvariables}
\end{table}

To determine the types of places on each street, we resort to Foursquare. We associate each place on Foursquare with the closest street and categorize it using the first-level categories: arts, college, food, nightlife, outdoors, residential, shopping, and travel. We choose the first level to avoid data sparsity.

To test the extent to which safety is associated with the presence of specific places, we build a linear model that predicts safety scores from the presence of first-level Foursquare categories. That is, a street's \emph{predicted} safety score is computed from the fraction of places on it that fall into the different  categories:

\vspace{-6mm}
\begin{multline*}
safety_i = \alpha + \beta_1 arts + \beta_2 college + \beta_3 food + \beta_4 nightlife + \\ \beta_5 outdoors + \beta_6 residential + \beta_7 shopping, +  \beta_8 travel + e.
\end{multline*}

It turns out that the regression shows an adjusted $R^2$  of 74\%, suggesting that safety can be accurately predicted only from the presence of Foursquare venues.  The corresponding beta coefficients (Table~\ref{table:predvariables}, column 3) suggest that safe streets tend to be associated with outdoor places (mainly parks), while unsafe ones with residential bits of central London that have no parks. This might appear surprising at first. However, further investigation shows that, in Central London, well-to-do residential areas are often associated with parks, while deprived areas are not. Therefore, this result can be explained by a strong interaction effect between residential streets and parks. 

\subsection{Research Question 4}
\emph{Can walkable streets be identified by the presence of specific types of places?}

Walkability and safety are related to each other. However, safe streets might not be necessarily walkable, and vice versa. In fact, the correlation between those two scores is as low as $r=0.22$. Having answered the questions about safety, it is now interesting to explore those about walkability.  

To test the extent to which walkability is associated with the presence of specific places, we regress a street's \emph{predicted} walkability score with the fraction of places on it that fall into the different  categories:
\begin{multline*}
walkability_i = \alpha + \beta_1 arts + \beta_2 college + \beta_3 food + \beta_4 nightlife + \\ \beta_5 outdoors + \beta_6 residential + \beta_7 shopping, +  \beta_8 travel + e.
\end{multline*}
We find that the above model has an adjusted $R^2$  of 33\%. That is, 33\% of the variability of the walkability score can be explained only by the presence of specific Foursquare venues. The beta coefficients of the model are shown in Table~\ref{table:predvariables} (column 2) and tell us that the presence of residential areas drives most of the predictive power of the regression.

\subsection{Research Question 5}
\emph{Can walkable streets be identified by walkability-related photo tags ?}

\begin{figure*}[t!] \begin{center}
\subfigure[Pearson correlations between walkability scores and \mbox{ } \break presence of car tags ($1^{st}$ bar), of  walkability tags ($2^{nd}$ bar),  \mbox{ } \break and of $z$-transformed combination of both ($3^{rd}$ bar). ]{\includegraphics[width=.45\textwidth]{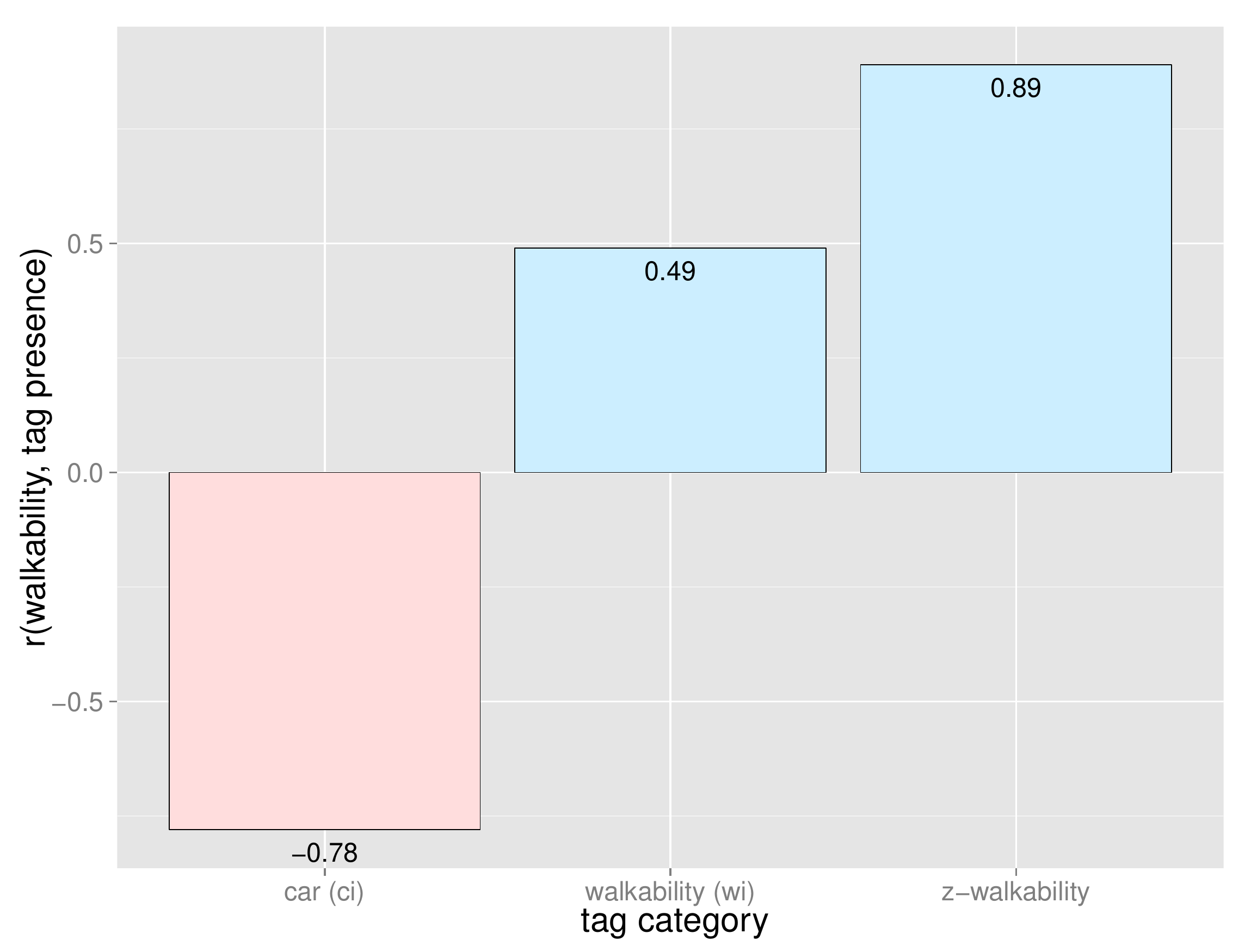} \label{fig:cor_tags_walkability}}
\subfigure[Pearson correlation coefficients between walkability scores and $z$-transformed presence of \emph{walkability tags} for increasing number of photos per segment. Whiskers represent the $2^{nd}$
and $98^{th}$ percentiles.]{\includegraphics[width=.45\textwidth]{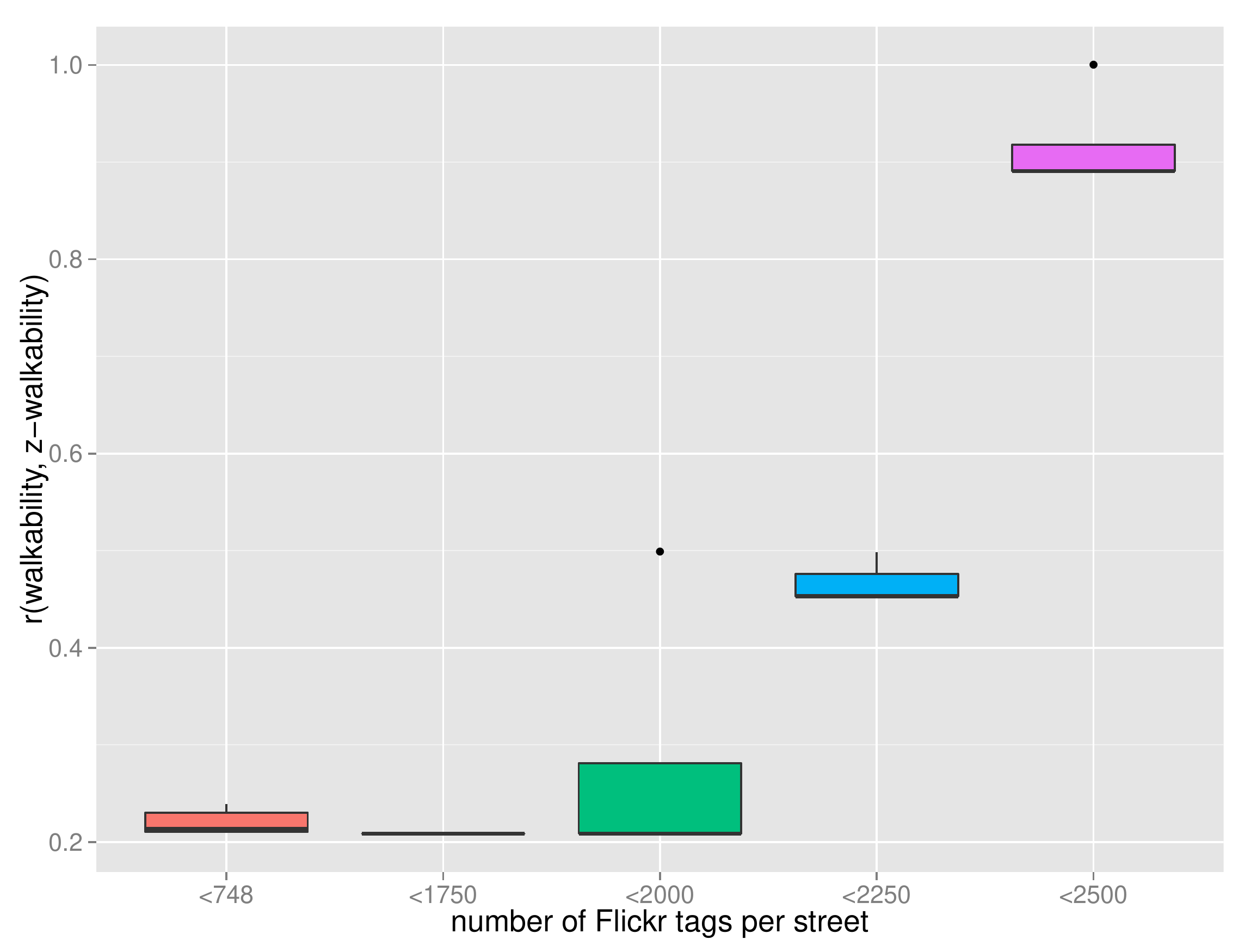} \label{fig:tags_walkability}}
\end{center} \vspace*{-2mm}
\caption{The digital life of walkable streets. Walkable streets can be identified by the presence of walkability-related picture tags (panel \emph{(a)}), but such an identification needs at least ~2.2K tags per street segment (panel \emph{(b)}). } 
\vspace*{-1mm}\label{} \end{figure*}

To build the list of keywords associated with the concept of walkability,  we hand-code relevant literature. We use the \emph{Grounded Theory} approach~\cite{Corbin:GT}, which is a systematic framework in the social sciences involving theory-driven content analysis that aims at identifying a set of words that best represent a certain concept. More specifically, we use \emph{line-by-line} coding. This generates a set of words conceptually associated with walkability in three steps:

\emph{1. Collecting documents.} The gold standard should cover the topic of walkability as comprehensively as possible. We collected a set of  documents that fall into three categories: 1) recent news articles from online media;  2) academic papers; and 3) recent reports from public organizations or governments. This collection includes: 6 news articles, 8 academic papers, and 2 reports. 

\emph{2. Annotating the documents.} Three annotators coded the list of keywords. The annotators separately read each document line-by-line and highlighted any word they felt to be related to walkability. We then combined their annotations to generate two distinct lists: one  \emph{merges} the three sets of annotations, and the other \emph{intersects} them.

\emph{3. Validating annotations.} To quantitatively validate the two lists, we measure \emph{agreement} among annotators defined as the ratio of the size of the merged word sets over the size of the intersected sets. The agreement is $84\%$, suggesting high agreements between the two lists. 

High agreement emerges because the words that characterize walkability are quite well recognizable as such by different people, and therefore we can safely use them to identify photos related to the walkability concept.  In our experiments, we adopt a very conservative approach and use the intersection list, which contains these terms: sidewalk, footway, street light, clean street, pedestrian, bench, resting, tree, greenery, art, architecture, historical, bike, private, hill, and social. One could informally see that those keywords indeed refer to the domain of walkability. However, those words by no means represent an exhaustive list and, as such, it is not clear whether we will observe any relationship between the presence of such keywords and walkability scores.

To balance those walkability tags (which mostly reflect positive associations), we create a  list containing the tag `car(s)'. That is because cars are often associated with poor walkability~\cite{alexander1977pattern}. Having a single-term list  might seem oversimplified. However, to appreciate the negative impact of cars on walkability, recall that for Jeff Speck's General Theory of Walkability, a walk has to satisfy four main conditions. It must be not only useful, comfortable, and interesting, but also safe~\cite{speck2012walkable}. By safe, he simply means that ``the street has been designed to give pedestrians a fighting chance against being hit by automobiles.'' In later chapters, he adds: ``Contrary to perceptions, the greatest threat to pedestrian safety is not crime, but the very real danger of automobiles moving quickly.'' In a similar way, Christopher Alexander notes: ``Cars give people wonderful freedom and increase their opportunities. But they also destroy the environment, to an extent so drastic that they kill all social life.''~\cite{alexander1977pattern} The effect of cars on health and social life is well documented: higher traffic exposure results into more heart attacks~\cite{gold13}, and hidden parking boosts retail sales, property values, appeal, and liveability~\cite{speck2012walkable,song07valuing}. The entire aesthetic capital of a neighborhood can be squandered by the sole presence of cars~\cite{quercia14aesthetic}.

Therefore, for each street segment $i$, we compute a $z$-transformed walkability score from Flickr tags:
\begin{equation*}
\zwalkability = \frac{w_i - \mu_w}{\sigma_w} - \frac{c_i - \mu_c}{\sigma_c},
\end{equation*}
where $w_i$ ($c_i$) is the fraction of tags that match our walkability-related keywords (match `car') on street segment $i$; $\mu_m$  ($\mu_o$) is the fraction of tags that match our walkability-related keywords (match `car'), averaged across all segments; and $\sigma_w$ ($\sigma_c$) is the corresponding standard deviation. 

Having those z-transformed scores, we can now correlate them with \emph{walkability}  (Figure~\ref{fig:cor_tags_walkability}). We find strong correlations between walkability and presence of tags mentioning cars: the correlation with $c_i$ is as high as -0.78. Given that the matching is done on a single term, this effect size is unexpectedly high, yet it speaks to the devastating effect of cars on walkability. As one expects, there is a positive correlation with the 
 walkability-related tags (i.e., the correlation with $w_i$ is 0.49). By then combining those two lists with the formula above, we obtain a correlation with $\zwalkability$ of 0.89.

However, those correlations might hold only for  data-rich streets. By binning  streets whose number of tags fall into the same range together, we find that the correlation between walkability score and $\zwalkability$ increases with the number of tags per segment and tends to become stable ($r > 0.85$) after collecting at least 2500 tags per street (Figure~\ref{fig:tags_walkability}). This translates into a considerable number of pictures required for attaining a reasonable prediction accuracy (of the order of hundreds).  That is likely because matching our keywords with Flickr tags yield sparse results. To partly fix that, in the future, one could either  enrich our list of keywords or use unsupervised techniques to learn the statistical associations of  Flickr tags with walkability score.

\section{Discussion}
\label{sec:discussion}

After successfully extracting social media metrics that reflect the walkability of physical streets, we now discuss a few open questions.

\subsection{Practical Implications}
We foresee many opportunities to practically apply walkability modeling, including:

\begin{description}
\item \emph{Room booking.} When tourists choose a place to stay, a system could make educated guesses about which places are walkable, and which are not. Walkability score might be a good indicator of whether they need to rent a car, for example.

\item \emph{Urban route recommendations.} In the near future, new way-finding tools might well suggest not only shortest routes but also short ones that are pleasant and walkable~\cite{quercia14shortest}. However, the data needed by those tools is available only for a limited number of cities and, when available,  is static. By contrast, our proposal allows for  timely recommendations of routes at scale.

\item \emph{Real-estate.} The use of walkability sites by real-estate agencies continues to grow~\cite{speck2012walkable}. With such a demand,  many municipalities are under pressure to collect relevant data and make it easily accessible.  For cash-strapped municipalities, it is usually difficult to obtain suitable data for computing walkability, owing to the high time, effort, and financial costs. Being based on social media mining, our approach promises to predict walkability scores at far lower cost.
\end{description}

\subsection{Theoretical Implications}
One contribution of this work to existing theory is the study of walkability dimensions as manifested in Flickr and Foursquare. As a result, we have ascertained the reliability of such sites for studying walkability. That is important, not least because it suggests that social media might offer unprecedented opportunities for theory. With real-time and fine-grained data, can we measure new indicators concerning walkability for which we have had no data (e.g., lifestyles and interests of individual street dwellers)? 

\subsection{Limitations}
Our approach is not able to  profile areas that have little Flickr activity\footnote{For Foursquare, activity is not required as the mere presence of venues suffice.}. Yet, it has two main advantages over the current state of the art: it adapts with time (unlike results from manual data collection efforts, that are costly to update), and it establishes smart defaults for places for which no census data is available. In the future, to design a system that works in a broader range of situations, one could augment our model with street design features, which have to be collected only once in while as they do not tend to massively change over time.

\section{Conclusion}
\label{sec:conclusion}
Our analysis has demonstrated that the relationship between behavioral features and walkability does not only hold in the offline world but also holds in the online world.  This provides evidence that users' offline communities have a noticeable effect on their online interactions. To appreciate the importance of this insight, consider the relationship between the types of streets people experience in their cities and the social media content they generate while being on those streets. We have tested this relationship for the first time and found that, indeed, Flickr uploads from dwellers of walkable streets differ from those of unwalkable ones,  mainly in terms of upload time and tagging. 

More broadly, our results suggest that it is possible to effectively profile the walkability of city streets from their dwellers' social media posts in an unobtrusive way. Many opportunities open up from here for designers and researchers alike. Mobile app designers, for example, may create new recommendation services that  combine walkability predictions with traditional mapping  tools. On the other hand, comforted by our validation work, urban researchers might well be enticed to  use social media  to answer theoretical questions that could not have been tackled before because of lack of data.

\section{Acknowledgments} We thank Elizabeth Daly for her timely feedbacks on the manuscript.

\balance

\bibliographystyle{abbrv}
\bibliography{references}

\begin{thebibliography}{10}

\bibitem{alexander1977pattern}
C.~Alexander, S.~Ishikawa, and M.~Silverstein.
\newblock {\em {A Pattern Language: Towns, Buildings, Constructions}}.
\newblock {Oxford University Press}, 1977.

\bibitem{angel68}
S.~Angel.
\newblock {The Ecology of Night Life}.
\newblock {\em {Center for Environmental Structure, Berkeley}}, 1968.

\bibitem{ball2012livable}
S.~Ball.
\newblock {\em {Livable Communities for Aging Populations: Urban Design for
  Longevity}}.
\newblock {Wiley}, 2012.

\bibitem{bentley2012}
F.~Bentley, H.~Cramer, W.~Hamilton, and S.~Basapur.
\newblock {Drawing the city: differing perceptions of the urban environment}.
\newblock In {\em {Proceedings of ACM Conference on Human Factors in Computing
  Systems (CHI)}}, 2012.

\bibitem{ramirez-indicators}
L.~K. Brennan~Ramirez.
\newblock {Indicators of Activity-Friendly Communities}.
\newblock {\em {American Journal of Preventive Medicine}}, 2006.

\bibitem{ceccato2012urban}
{Ceccato, V.}
\newblock {\em The Urban Fabric of Crime and Fear}.
\newblock {Springer}, 2012.

\bibitem{Cheng:EtAl:11}
Z.~Cheng, J.~Caverlee, and K.~Lee.
\newblock Exploring millions of footprints in location sharing services.
\newblock In {\em {Proceedings of the AAAI International Conference on Webblogs
  and Social Media (ICWSM)}}, 2011.

\bibitem{Corbin:GT}
J.~M. Corbin.
\newblock {\em {Basics of Qualitative Research: Techniques and Procedures for
  Developing Grounded Theory}}.
\newblock Sage, 2008.

\bibitem{cortright09walking}
J.~Cortright.
\newblock {Walking the Walk: How Walkability Raises Home Values in U.S.
  Cities}.
\newblock {\em {CEOs for Cities}}, 2009.

\bibitem{cramer11performing}
H.~Cramer, M.~Rost, and L.~E. Holmquist.
\newblock {Performing a check-in: emerging practices, norms and 'conflicts' in
  location-sharing using foursquare}.
\newblock In {\em {Proceedings of ACM Conference on Human Computer Interaction
  with Mobile Devices and Services (MobileHCI)}}, 2011.

\bibitem{crandall09mapping}
D.~J. Crandall, L.~Backstrom, D.~Huttenlocher, and J.~Kleinberg.
\newblock Mapping the world's photos.
\newblock In {\em {Proceedings of ACM Conference on World Wide Web (WWW) }},
  2009.

\bibitem{cranshaw12livehoods}
J.~Cranshaw, R.~Schwartz, J.~Hong, and N.~Sadeh.
\newblock {The Livehoods Project: Utilizing Social Media to Understand the
  Dynamics of a City}.
\newblock In {\em {International AAAI Conference on Weblogs and Social Media
  (ICWSM)}}, 2012.

\bibitem{cranshaw12}
J.~Cranshaw, R.~Schwartz, J.~I. Hong, and N.~Sadeh.
\newblock {The Livehoods Project: utilizing social media to understand the
  dynamics of a city}.
\newblock In {\em Proc. of ICWSM}. AAAI, 2012.

\bibitem{gold13}
J.~M.~S. Diane R.~Gold.
\newblock {Air Pollution, Climate, and Heart Disease}.
\newblock {\em {Circulation}}, 2013.

\bibitem{eagle10}
N.~Eagle, M.~Macy, and R.~Claxton.
\newblock {Network diversity and economic development}.
\newblock {\em Science}, 328:1029--1031, 2010.

\bibitem{elvidge97}
C.~D. Elvidge, K.~E. Baugh, E.~A. Kihn, H.~W. Kroehl, and E.~R. Davis.
\newblock {Mapping city lights with nighttime data from the DMSP Operational
  Linescan System}.
\newblock {\em {Photogrammetric engineering and Remote Sensing}}, 1997.

\bibitem{elvidge01}
C.~D. Elvidge, M.~L. Imhoff, K.~E. Baugh, V.~R. Hobson, I.~Nelsonc, J.~Safran,
  J.~B. Dietz, and B.~T. Tuttle.
\newblock {Night-time lights of the world: 1994-1995}.
\newblock {\em {Photogrammetry and Remote Sensing}}, 2001.

\bibitem{hess2008introduction}
K.~Hess.
\newblock {\em {Introduction to Private Security}}.
\newblock {Cengage Learning}, 2008.

\bibitem{jacobs1993great}
A.~Jacobs.
\newblock {\em {Great Streets}}.
\newblock {MIT Press}, 1993.

\bibitem{jacobs1961death}
J.~Jacobs.
\newblock {\em {The Death and Life of Great American Cities}}.
\newblock {Random House}, 1961.

\bibitem{kramer10unobtrusive}
A.~D. Kramer.
\newblock {An Unobtrusive Behavioral Model of ``Gross National Happiness''}.
\newblock In {\em {Proceedings of ACM Conference on Human Factors in Computing
  Systems (CHI)}}, 2010.

\bibitem{depriv2010}
M.~Lad.
\newblock English indices of deprivation 2010.
\newblock Technical report, {Department for Communities and Local Government},
  2011.

\bibitem{lee2011reversing}
{Lee, R.E. and McAlexander, K. and Banda, J.}
\newblock {\em {Reversing the Obesogenic Environment}}.
\newblock {Human Kinetics}, 2011.

\bibitem{leinberger12coveted}
C.~B. Leinberger.
\newblock {Now Coveted: A Walkable, Convenient Place}.
\newblock {\em {New York Times}}, 2012.

\bibitem{lindqvist11mayor}
J.~Lindqvist, J.~Cranshaw, J.~Wiese, J.~Hong, and J.~Zimmerman.
\newblock {I'm the mayor of my house: examining why people use foursquare - a
  social-driven location sharing application}.
\newblock In {\em {Proceedings of ACM Conference on Human Factors in Computing
  Systems (CHI)}}, 2011.

\bibitem{lynch1960}
K.~Lynch.
\newblock {\em The Image of the City}.
\newblock Urban Studies. {MIT Press}, 1960.

\bibitem{mao13}
H.~Mao, X.~Shuai, Y.~Y. Ahn, and J.~Bollen.
\newblock {Mobile communications reveal the regional economy in Cote d'Ivoire}.
\newblock In {\em Proc. of NetMob}, 2013.

\bibitem{Methorst2010}
R.~Methorst, H.~Monterde, R.~Risser, D.~Sauter, M.~Tight, and J.~Walker.
\newblock {Pedestrians' Quality Needs}.
\newblock {\em European Cooperation in Science and Technology (COST)}, 2010.

\bibitem{tassos12}
A.~Noulas, S.~Scellato, R.~Lambiotte, M.~Pontil, and C.~Mascolo.
\newblock {A Tale of Many Cities: Universal Patterns in Human Urban Mobility}.
\newblock {\em {PLoS ONE}}, 2012.

\bibitem{quercia12a}
D.~Quercia, J.~Ellis, L.~Capra, and J.~Crowcroft.
\newblock {Tracking Gross Community Happiness from Tweets}.
\newblock In {\em {Proc. of ACM Conference on Computer Supported Cooperative
  Work (CSCW)}}, 2012.

\bibitem{quercia14aesthetic}
D.~Quercia, N.~Ohare, and H.~Cramer.
\newblock {Aesthetic Capital: What Makes London Look Beautiful, Quiet, and
  Happy?}
\newblock In {\em {Proc. of ACM Conference on Computer Supported Cooperative
  Work (CSCW)}}, 2014.

\bibitem{quercia13maps}
D.~Quercia, J.~P. Pesce, V.~Almeida, and J.~Crowcroft.
\newblock {Psychological Maps 2.0: A web gamification enterprise starting in
  London}.
\newblock In {\em {Proceedings of ACM Conference on World Wide Web (WWW) }},
  2013.

\bibitem{quercia14shortest}
D.~Quercia, R.~Schifanella, and L.~M. Aiello.
\newblock {The Shortest Path to Happiness: Recommending Beautiful, Quiet, and
  Happy Routes in the City}.
\newblock In {\em Proceedings of the 25th ACM Conference on Hypertext and
  Social Media (HT)}, 2014.

\bibitem{quercia12b}
D.~Quercia, D.~O. S\'eaghdha, and J.~Crowcroft.
\newblock Talk of the city: our tweets, our community happiness.
\newblock In {\em {Proc. of International AAAI Conference on Weblogs and Social
  Media (ICWSM)}}. AAAI, 2012.

\bibitem{sampson12when}
R.~J. Sampson.
\newblock {When Things Aren{\textquoteright}t What They Seem: Context and
  Cognition in Appearance-Based Regulation}.
\newblock {\em {Harvard Law Review Forum}}, May 2012.

\bibitem{smith13b}
C.~Smith, D.~Quercia, and L.~Capra.
\newblock Finger on the pulse: identifying deprivation using transit flow
  analysis.
\newblock In {\em {Proc. of ACM ACM Conference on Computer Supported
  Cooperative Work (CSCW)}}, pages 683--692, 2013.

\bibitem{speck2012walkable}
J.~Speck.
\newblock {\em {Walkable City: How Downtown Can Save America, One Step at a
  Time}}.
\newblock {Farrar, Straus and Giroux}, 2012.

\bibitem{Gleave14}
{Steer Davies Gleave}.
\newblock {Legible London}.
\newblock {\em {TfL Research Report}}, 2014.

\bibitem{thomee14time}
B.~Thomee, J.~G. Moreno, and D.~A. Shamma.
\newblock {Whose Time Is It Anyway? Investigating the Accuracy of Camera
  Timestamps}.
\newblock In {\em {Proceedings of ACM Conference on Multimedia (MM)}}, 2014.

\bibitem{transport2012}
{Transport for London}.
\newblock {Attitudes towards Walking}.
\newblock {\em {TfL Research Report}}, March 2012.

\bibitem{traunmueller14mining}
M.~Traunmueller, G.~Quattrone, and L.~Capra.
\newblock {Mining Mobile Phone Data to Investigate Urban Crime Theories at
  Scale}.
\newblock In {\em {$6^{th}$ International Social Computing Conference
  (SocInfo)}}, 2014.

\bibitem{transp2013}
{UK Department for Transport}.
\newblock Road accidents and safety statistics.
\newblock Technical report, 2013.

\bibitem{geological10}
{US Geological Survey}.
\newblock {Shuttle Radar Topography Mission}.
\newblock 2010.

\bibitem{song07valuing}
J.~S. Yan~Song.
\newblock {Valuing spatial accessibility to retailing}.
\newblock {\em {Journal of Retailing and Consumer Services}}, 2007.

\bibitem{Quercia2014lightweight}
V.~Zambaldi, J.~P. Pesce, D.~Quercia, and V.~Almeida.
\newblock {Lightweight Contextual Ranking of City Pictures: Urban Sociology to
  the Rescue}.
\newblock In {\em {International AAAI Conference on Weblogs and Social Media
  (ICWSM)}}, 2014.

\end{thebibliography}
\end{document}